# Radiation damage to neutron and proton irradiated GaAs particle detectors

M.Rogalla* , Th.Eich, N.Evans, S. Joost, M. Kienzle, R.Geppert, R.Göppert, R.Irsigler, J.Ludwig, K.Runge, Th.Schmid
Albert-Ludwigs-Universität Freiburg, Fakultät für Physik, Hermann-Herder-Str. 3, 79104 Freiburg, Germany

*As part of the RD8 collaboration*

**Abstract**

The radiation damage in 200 µm thick Schottky diodes made on semi-insulating (SI) undoped GaAs Liquid Encapsulated Czochralski (LEC) bulk material with resistivities between 0.4 and $8.9*10^{7}$ Ωcm were studied using α-spectroscopy, signal response to minimum ionising particles (MIPs), I-V and CV-measurements. The results have been analysed to investigate the influence of the substrate resistivity on the detector performance after neutron and proton irradiation. The leakage current density, signal response to α-particles and MIPs show a strong dependence on the resistivity before and after irradiation. An observed decrease of the electron mean free drift length before and after irradiation with increasing substrate resistivity can be explained by a model involving the different ionisation ratios of defects, which are introduced by the irradiation. Comparison of the radiation damage due to neutrons and protons gives a hardness factor of $7 \pm 0.9$ for 24 GeV/c protons. The best detectors show a response to MIPs of 5250 $e^-$ at 200 V reverse bias after a irradiation level of $2*10^{14}$ p/cm².

* Corresponding author. Tel. +49 761 2035911, fax +49 761 2035931, e-mail rogalla@ruhp8.physik.uni-freiburg.de

## Introduction

Schottky diodes made from commercial undoped, semi-insulating (SI) Gallium Arsenide (GaAs) materials have been shown to work well as radiation detectors. This material offers high charge carriers mobilities, good signal-to-noise ratio and a good detection efficiency for minimum ionizing particles (MIPs) [1,2]. For its use as an inner tracking detector at future high energy colliders, such as the LHC (Large Hadron Collider) at CERN, supreme radiation resistance is necessary. Experiments have shown a good detector performance of GaAs Schottky diodes after irradiation with high energetic neutrons fluences up to $4*10^{14}$ n/cm$^2$ [3]. In the ATLAS inner tracking detector area there are additional high fluences of high energetic protons and pions in the range of $10^{13}$ cm$^{-2}$ yr$^{-1}$ each [4]. Therefore the magnitude of signals obtained from MIPs, mean free drift length of charge carriers and the leakage current density of various GaAs particle detectors after irradiation with neutrons (ISIS) and 23 GeV protons (CERN) are compared. We also investigate the radiation hardness of different types of substrate materials and determine the damage function of each material for the detectors parameters.

## 1. Detectors and Radiation Sources

The detectors were Schottky diodes fabricated on Liquid Encapsulated Czochralski (LEC) grown semi-insulating GaAs obtained from different manufactures. The resistivity of the substrates ranges between 0.4 and $8.9*10^7$ $\Omega$cm at 20°C (Table 1). All detectors have a thickness of 200 µm and are made out of 3" wafers. The ohmic contact covers the whole back side of the detectors, whereas the front side was patterned with circular Schottky contacts of 2 and 3 mm in diameter. Before the evaporation of the contacts, the wafers were dipped for 30 seconds in $HCl:H_2O$ (1:1) and $NH_4OH:H_2O_2:H_2O$ (5:1:50) to reduce the contamination of the surface with native oxides. For the back contact we used an alloyed ohmic Ni/Ge/Au multi layer or a Schottky contact biased in forward direction. As Schottky contacts we used Ti/Pt/Au, Ti/Ti-W or TiN/Ti-W layers. Some diodes feature guardrings at a distance of 50 μm from the diodes to avoid surface currents. The investigated detectors and their preparation are summarised in Table 1.

For this work the pad detectors were irradiated at the ISIS spallation neutron source of the Rutherford Appleton Laboratory. The average flux during the irradiation varied between $2.1*10^7$ and $2.4*10^8$ neutrons $cm^{-2}\ s^{-1}$ and the estimated uncertainty of the fluence is ± 17 %. The energy spectrum of the ISIS neutrons ranged from thermal energies up to $10^3$ MeV with a peak at 1MeV. According to the non ionizing energy loss (NIEL) hypothesis [5], where the bulk damage effects scales linearly with the total NIEL, we calculated the hardness factor for the neutrons delivered by ISIS using the spectra[1] given by [6] and the NIEL data of Ougouag et al. [7] assuming a similar saturation of the NIEL for energies higher than 20 MeV as in silicon [8]. This estimation gives a hardness factor of 2 ± 0.2, which means that the equivalent 1 MeV neutron fluence (NIEL(1 MeV,n) = 3.1 keV/cm) is a factor 2 higher then the measured fluences.

As the source of protons the Proton Synchrotron (PS) at CERN was used. The beam energy was 23 GeV with negligible spread. The average flux during an entire irradiation was approximately $3*10^9$ protons $cm^{-2}\ s^{-1}$. The detectors were irradiated up to a maximum fluence of about $2*10^{14}\ cm^{-2}$. The measurement of activation of aluminium foils provided dosimetry to an accuracy of about ± 3 % [9].

[1] only the fluence for neutrons with energy higher than 10 keV.

All detectors parameters were measured before irradiation and showed a homogeneity better than 5% for each material. After irradiation the homogeneity was of the same order and also better than 5%. As a result, for all parameters, the mean value is given. Up to now, all GaAs samples have shown no evidence of either self or reverse annealing. Therefore the data are given without any correction and was measured within a period of a month after irradiation.

## 2. Radiation damage from neutrons

### *2.1 Leakage current density and electrical properties*

Measurements of leakage current densities have been made as a function of reverse bias voltage for the different neutron fluences. The measurements have been taken in a temperature controlled setup using a Keithley 487 picoammeter/voltage source. Figure 1 shows as an example the I-V curves from diodes of wafer FR41 taken at 30°C before and after irradiation with neutrons from ISIS. The I-V characteristic has two distinct regions: a) a region in which the leakage current increases slowly with reverse bias voltage (saturation region) and b) a region where the current increases faster with applied voltage (breakdown region). Before irradiation there is a sudden breakdown at about 140 V. After irradiation the leakage current increases linearly for bias voltages higher then the former breakdown voltage. The slope of the current voltage characteristics in this soft breakdown region decreases with increasing neutron fluence. The breakdown voltage (defined as the intersection point of a linear fit on the saturation region and breakdown region), decreases with increasing fluence.

To compare the I-V characteristics of the detectors made from different materials we show in figure 2 the leakage current density at 100 V (30°C) as a function of the neutron fluence. For all detectors the leakage current densities increase with radiation dose. Before irradiation the detectors on wafer FR76 with the highest resistivity of $5.9 \times 10^7$ Ωcm show the smallest leakage current density of 34 nA/mm² (30°C) and at the highest fluence of $1.72*10^{14}$ n/cm² a value of 67.2 nA, only about twice as high. To investigate the change in the leakage current density, we analyse the I-V characteristic in forward direction using a Norde plot and determine the Schottky barrier height as a function of irradiation level at 20°C [10]. With irradiation of all materials, the Schottky barrier height shows a slight decrease of about 0.04 eV, which explains qualitative the increase of the leakage current (see for example for the diodes on wafer FR41 in figure 3). As mentioned in a previous paper [14] concerning Schottky diodes made of SI-GaAs the Schottky barrier height depends on the Fermi-level position in the bulk material, so the change of the leakage current is not due to surface damage effects, but more likely due to the bulk damage.

The breakdown voltage decreases with increasing neutron fluence e.g. from about 210 V down to 150 V (fig. 4) for the high resistivity material FR76. The detectors from wafer FR41 show the smallest change of about 20 V. From the change of breakdown voltage we conclude that the damage due to neutron irradiation decreases voltage necessary to activate the whole detector depth.

To verify this decrease in space charge density we perform CV-analysis of the barriers created by the Schottky contact. The net ionized defect density $N_{net}^{+}$ in the space charge region is given by

$$N_{net}^{+} = \left(\frac{2}{q\varepsilon S^2}\right)\left(\frac{\partial V_B}{\partial C^2}\right) \quad , \tag{1}$$

where $\varepsilon$ is the semiconductor permittivity, S the area of the diodes, $V_B$ the bias voltage and C the measured capacitance. For the CV-measurements we used an impedance/gain-phase analyser (Schlumberger SI1600). Because of the low emission rates of the deep levels in SI-GaAs [11], a test signal with a frequency of 10 Hz (test signal amplitude 0.05 V, room temperature) was applied. Figure 5 shows as an example the $1/C^2$-V plot of a diode from wafer

MCP90 before and after the highest neutron fluence. The slope of the curves increases with bias voltage and irradiation level, which implies a decrease of the space charge concentration (Figure 6). The decrease of the space charge density, within bias voltages between 7.5 and 35 V of about two orders of magnitude is typical for Schottky diodes made on SI-GaAs, and was explained by a shift of the quasi-Fermi-level due to increase of the concentration of electrons in the conduction band caused by the leakage current [12,13,14]. Figure 7 summarises the space charge density as a function of neutron fluence at 20 V bias for different materials. With this observed decrease of the space charge density with increasing neutron fluence we can explain the shift of the breakdown voltage towards smaller values. The largest variation of the space charge density at 20 V bias shows the diodes labelled MCP90 with a value of about $6*10^{14}$ $cm^{-3}$ before and $1.5*10^{14}$ $cm^{-3}$ after highest neutron fluence.

*2.2 Detector performance*

To investigate the degradation of the detector performance due to neutron irradiation, we measured the charge collection efficiency (CCE) for α-particles and the signal height for minimum ionizing particles (MIPs).

The CCE was measured twice for α-particles from an $^{241}$Am source. The detector consists of a Schottky diode on the front side and an ohmic contact on the reverse side. Both sides were subsequently irradiated with Alpha particles. The spectroscopy chain includes an Vitrom (559-064) charge sensitive preamplifier followed by an ORTEC 579 amplifier-shaper with gaussian shaping and a time constant of 500 ns. The penetration depth of the α-particles within GaAs is less than 20 µm. For backside exposure the signal is primarily due to the movement of holes. In any case a high electric field is necessary for obtaining a signal. And can only be seen for full active detectors.

Figure 8 shows the CCE as a function of reverse bias voltage before and after neutron irradiation of diodes from wafer MCP90. The solid lines represents the values for the front side exposure, where the signal is primarily due to the electron drift in the detector. Before irradiation the CCE is proportional to the square root of applied voltage and saturates at a value of 51 %. It was not possible to measure the CCE for backside irradiation, because of the fast breakdown of the unirradiated detector. For all neutron fluences the CCE for α-particles incident from the Schottky, contact increases rapidly to a maximum, decreases slightly with higher bias voltage and grows slowly again for voltages higher than 300 V. With increasing fluence, the first maximum of the CCE moves to lower voltage. For the highest neutron fluence the CCE is almost independent of the bias voltage for values higher than 100 V. The CCE from the backside exposure (dotted lines) increases linearly with the bias voltages and the slope decreases with neutron fluence. The corresponding voltage, where the electrical field reaches through to the backside of the detector and a signal can be seen is in a good agreement with the measured breakdown voltage and decreases with increasing radiation damage.

According to Ramo´s theorem [15] the CCE is given by

$$CCE = \frac{\lambda_n}{d}\left(1-\exp\left(-\frac{d-x_0}{\lambda_n}\right)\right) + \frac{\lambda_h}{d}\left(1-\exp\left(-\frac{x_0}{\lambda_h}\right)\right) \quad (2)$$

where $\lambda_n$ and $\lambda_h$ denote the mean free drift lengths of electrons and holes, d the detector thickness and $x_0$ represents the generation point of electron hole pairs. To determine the mean free drift length from the α-spectra, we assume that $\lambda_{n/h}$ is independent of the position in the detector and consider the energy loss distribution of the α-particles calculated with Trim [16]. In any case the mean free drift length is, at bias voltages higher than the full active voltage, approximately constant for electrons due to the saturation of the drift velocity for electric fields higher than $10^4$ V/cm. This explains the saturation of the CCE of electrons. For holes this is not such a good approximation, because of the linear increase of the CCE observed at low neutron fluences. Despite of this the determined mean free drift length is approximately the mean value. For the highest irradiation level (where the CCE reaches a plateau, indicating a saturation of the drift length) the assumption is valid. Figure 9 shows the mean free drift length of electrons and holes (20 °C, 300 V bias) as a function of neutron fluence for all three materials. Before irradiation we used the maximum of the CCE for electrons to calculate the mean free drift length. For diodes from wafer MCP90 we observed a continuos decrease of the drift length for both carrier types, down to 55 µm for electrons and 26 µm for holes. The diodes from wafer FR41 and FR76 showed a slight increase of the mean free drift length for electrons. For all fluences, regardless of the material, the mean free drift length is smaller for holes than for electrons.

The detector response to MIPs was measured using a $^{90}$Sr source and a Si-diode trigger to select relativistic beta particles. Spectra were taken at different reverse bias voltages between 25 and 400 V. Because relativistic β-particles pass through the detector, electron-hole pairs are generated along the path through the detector and hence it is not possible to separate the signal due to holes from that due to electrons. But we can conclude from the α-spectra (i.e. that the mean free drift length in all irradiated diodes is for holes smaller than for electrons) that the signal for MIPs is primarily due to electrons. For example, the signal heights (most probable value) for the diodes of wafer MCP90, obtained with a shaping time of 500 ns before and after exposure to different neutron fluences, are plotted in Figure 10. Before irradiation the signal height shows the typical approximate linear dependence on reverse bias voltage. For all radiation levels, the detector shows a similar signal dependence on reverse bias voltage: first a linear rise for small bias voltages then a sudden decrease of the slope, or even a saturation. The voltage at which the change of slope occurs decreases with increasing neutron fluence and coincides with the breakdown voltage.

Figure 11 summarises the signal pulse height data of MIPs for all neutron irradiated materials at a bias voltage of 200 V. Before and after the lowest fluence it was not possible to measure the signal height for the detectors from wafer FR41 and FR76, because of the fast breakdown. For the detectors from wafer MCP90 the signal height decreases from 15000 $e^-$ down to about 7700 $e^-$. With the detectors from wafer FR41 we observe the signal of 8000 $e^-$ at the maximum neutron fluence.

## 3. Radiation damage of protons

### *3.1 Leakage current density and electrical properties*

Figure 12 and 13 shows an example of I-V-characteristics of diodes from wafer MCP90 and OUT44 for various proton fluences up to about $2*10^{14}$ $cm^{-2}$. The characteristics was measured at 30°C for reverse bias voltages up to 400 V. The leakage current density of the diodes from wafer MCP90 (with a resistivity of $4.9*10^7$ Ωcm at 20°C) shows an increase of the leakage current density, similar to the neutron irradiated materials. The I-V-characteristic has also distinct regions of saturation and soft breakdown, as for the neutron irradiated diodes. We observed a fast change of leakage current density between the non irradiated and proton irradiated diodes with the lowest fluence of a factor 2 in the saturation region. At a reverse bias voltage of 100 V the leakage current density increases slowly with further irradiation up to 76.7 $nA/mm^2$ at the maximum proton fluence. The diodes from wafer OUT44 (with a resistivity of $2.1*10^7$ Ωcm at 20°C) show a totally different I-V characteristic (fig. 13). There is still a saturation and soft breakdown region, but the leakage current density decreases with increasing proton fluence from about 300 $nA/mm^2$ before irradiation down to 70 $nA/mm^2$ at a fluence.

Figure 14 summarises the leakage current densities of all irradiated detectors at a bias voltage of 100 V and a temperature of 20°C. For substrates with a resistivity between $4.8*10^7$ (FR41) and $8.9*10^7$ Ωcm (FR104) the leakage current density increases with proton fluence. The leakage current density of the diodes made of low ohmic semi-insulating materials like FR12 ($0.42*10^6$ Ωcm) and OUT44 ($2.1*10^7$ Ωcm) shows a fast decrease with increasing proton fluence. The leakage current density of the diodes from wafer OUT50 ($2.4*10^7$ Ωcm) only slightly decreases from 33 nA/mm² down to 30 nA/mm² at a lowest fluence and is for further increase of proton fluence approximately constant. For all diodes except from wafer FR12 and FR104 the leakage current density of about 30 nA/mm² at a proton fluence of about $2*10^{14}$ p/cm² is independent of the value before irradiation. For material FR12 with the lowest resistivity a leakage current density of 41 nA/mm² and for FR104 with the highest resistivity a value of 20.5 nA/mm² was measured after a proton fluence of $2.9*10^{14}$ $cm^{-2}$.

From the given data we conclude that the leakage current density as function of proton fluence depends on the substrate resistivity before irradiation. But for all materials the leakage current density is small in comparison to irradiated Si-detectors [17] and allows operation at room temperature. Also a bias voltage larger then the full active voltage can be applied due to the softer breakdown after irradiation. The breakdown voltage is given in figure 15 for all materials before and after irradiation and shows quite different behaviour for the individual material types. For all materials except for the low ohmic FR12 and OUT44 the breakdown voltage decreases with increasing proton fluence. The biggest decrease in breakdown voltage was measured for the diodes from wafer FR104 from 270 V down to 130 V at the highest proton fluence. The smallest change occurs for the diodes made of wafer OUT50 and is only about 20 V. For the low ohmic materials like e.g. FR12 the breakdown voltage increases rapidly from about 70 V to 270 V and then slight decreases down to 230 V with further irradiation.

The Schottky barrier height determined with the Norde plot is given for the diodes with ohmic contacts on the back side (fig. 16). All diodes show a decrease of Schottky barrier height similar to the neutron irradiated detectors. With this decrease of barrier height for the diodes from wafer FR41 and MCP90 we can explain qualitatively the increase of leakage current density. On the other hand, the decrease of leakage current density with proton fluence of the diodes from wafer OUT44 can not be explained with the change in Schottky barrier height only.

Another parameter, which can be determined from the Norde plot [10] is the resistivity of the substrate material. There are two mechanisms which can influence the resistivity during irradiation: first a change of the Fermi-level $E_f$ and second a change of the charge carrier mobilities. The change of the Fermi-level position with irradiation damage is due to the introduction of deep levels in GaAs and is well known in resistors made of GaAs [18]. Also a decrease of the hall mobilities with increasing neutron fluence was observed by Dubecky et al. [19]. All materials show the expected increase of the resistivity determined at 20°C with irradiation level (fig. 17). If we assume that after irradiation with protons the leakage current density is still a function of the Fermi-level position and rises with decreasing distance to the conduction band (as observed in unirradiated samples [14]) we can explain the largest increase of the resistivity for the material OUT44 from $2.1*10^7$ to $5.0*10^7$ Ωcm. In the material OUT44 the Fermi-level position shifts towards the middle of the band gap and causes a carrier removal in the conduction band, which results in an additional increase of resistivity, because of the larger electron than hole mobility in GaAs. Together with the decrease of charge carrier mobility with irradiation which can be assumed to be independent of the material type this will result in the largest increase of the resistivity with increasing proton fluence in comparison to the materials FR41 and MCP90. For these two materials the small measured increase of leakage current density indicated a small shift of the Fermi-level $E_f$ towards the conduction band, but the decrease in charge carrier mobilities means there is still an increase in resistivity.

To analyse the influence of the proton irradiation on the space charge density $N_{net}$, we perform in analogy to the neutron irradiated samples C-V-measurements at 10 Hz. Figure 18 shows the $N_{net}$ determined from the $1/C^2$-V plot of the diodes from wafer FR76 as a function of reverse bias voltage with proton fluence as parameter. Analogous to the observed behaviour of the neutron irradiated diodes, $N_{net}$ decreases with increasing reverse bias voltage and proton fluence. The $N_{net}$ measured at 37.5 V bias decreases with proton fluence from $9*10^{12}$ $cm^{-3}$ to $2.5*10^{11}$ $cm^{-3}$.

Figure 19 gives the $N_{net}$ at 20 V reverse bias at room temperature for all measured diodes. The $N_{net}$ decreases for all detectors by about an order of magnitude, in comparison to the maximum change in neutron irradiated material of a factor 4. Although the $N_{net}$ of diodes from OUT44 decreases with proton fluence down to values of about $2*10^{11}$ $cm^{-3}$ we have observed an increase in the breakdown voltage. This effect will be discussed in the following paragraph.

### *3.2 Detector performance*

In a previous paper [20] it was shown that the CCE for $\alpha$-particles depends on the Fermi-level position in the bulk material, or rather, on the substrate resistivity. The decrease of the CCE for electrons with increasing resistivity was explained as being due to the higher ionisation of the EL2-defect, which is an effective electron trap in the high field region. High ohmic material showed a better hole collection [21]. Therefore we analyse the CCE for both types of materials as a function of proton fluence.

Figure 20 shows as an example the CCE of $\alpha$-particles ($^{241}$Am) for detectors made of wafer OUT44 versus the reverse bias voltage and proton fluence. The solid lines give the value for front side and the doted line for back side $\alpha$-particles exposure. The CCE for front side exposure, which is mainly due the electron drift, shows before irradiation a fast increase up to 75 % at 100 V reverse bias. After the first irradiation step, the CCE for electrons increases up to an maximum, and saturates for higher bias voltages, similar in behaviour to the neutron irradiated samples. At the highest proton fluence the CCE has a value of about 20 % and is almost independent of the applied reverse bias voltage. The maximum CCE of electrons also shifts to lower bias voltages with increasing proton fluence, again as in the neutron irradiated detectors. Before irradiation it was not possible to detect for these detectors the CCE due to back side exposure, because of the fast breakdown at 130 V. After proton irradiation the CCE of holes can be measured for voltages higher than 140 V and shows a slight increase with reverse bias voltage. From the occurrence of signals from the back side exposure at bias voltages of 135 V (about 100 V lower than the breakdown voltage determined from the I-V characteristic) we can conclude that for low ohmic semi-insulating materials the breakdown voltage exceeds the full active voltage. This is also in good agreement with the observed decrease of net space charge density $N_{net}$ (fig.18).

There are two extreme cases however: a) low ohmic materials with good electron and poor hole collection (like FR12) and b) high ohmic materials like FR104 with good hole collection and small signals due to electrons. For these two materials the mean free drift length is given in figure 21 as a function of proton fluence. In unirradiated detectors from the low ohmic wafer we have measured an electron mean free drift length of about 4530 µm. An increase of the proton fluence results in a rapid decrease in the mean free drift length down to values of 41 µm at $2.9 \cdot 10^{14}$ p/cm². In comparison to these, the mean free drift length in the high ohmic material FR104 decreases only slow from 88 µm down to 28 µm. For holes the mean free drift length in the low ohmic material decreases only by a factor 2 down to 20 µm. The mean free drift length of holes in the material FR104 starts with a value of 760 µm before irradiation and decreases rapidly down to a slightly higher value of 22 µm than that for FR12. Because of the intersection of the mean free drift lengths for holes and electrons in the diodes from wafer FR104 at a proton fluence of about $2.5*10^{13}$ $cm^{-2}$, mean free drift length of holes is always smaller than for electrons after high proton fluences.

The dependence of the mean free drift length on the substrate resistivity before irradiation is given in figure 22 and 23 at two different radiation levels. At the low proton fluence of $5*10^{13}$ $cm^{-2}$, the mean free drift length of electrons shows a decrease from 118 µm for the low ohmic material FR12 down to 43 µm for the material with the highest resistivity of $8.9*10^{7}$ Ωcm. In comparison the mean free drift length of holes increases slightly from 26 to 38 µm with increasing resistivity. The mean free drift length of both carriers shows only a small variation at a proton fluence of $2*10^{14}$ $cm^{-2}$ (fig. 23). For holes the values are almost independent of the material type and the mean free drift length of electrons varies only by a factor 1.5 from 41 to 27 µm with increasing resistivity.

The decrease of mean free drift length as a function of resistivity at different proton fluences can be explained by the Fermi-level positions in the materials, which results in a different ionisation ratio of the deep levels. As shown for example by Pons and Bourgin [22] there is an introduction of shallow and deep levels in GaAs though irradiation with high energetic neutrons. Because of the fact that the radiation damage in semi-conductors is due to the displacement from the non ionising energy loss (NIEL) of the primary recoil atom [23], we expect that the high energetic neutrons and protons produce the same type of defects. The only difference is the primary process of the production of recoil atoms and fragments.

Many of the induced shallow defects have annealing temperatures lower than 250 K and are not present after irradiation at room temperature. Even so, if shallow levels are present after irradiation, there will be no signal loss, because of the fast electron/hole detrapping due to the Pool-Frenkel effect in the high field region [20]. The deep levels like the gallium antisite and arsenic antisite associated to the so called EL2 defect require annealing temperatures higher than 400 K and are therefore stable at room temperature. If we assume that the introduction rate of the deep defects is independent of the GaAs substrate, the only difference that occurs will be the ionisation ratio of the defects. For example the electron lifetime $\tau_e$ is given by the Schottky-Read-Hall statistics as

$$\tau_e = \frac{1}{\sigma N_d^+ v_{th}} \qquad (3)$$

where $N_d^+$ is the ionised donor concentration, $\sigma$ the capture cross section and $v_{th}$ the thermionic velocity of the electrons. In the low ohmic material the concentration of ionised deep levels e.g. the EL2, which acts as an effective electron trap in the high field region [20,24], is smaller with respect to high ohmic materials, due to the different $E_f$ position. This explains qualitatively the decrease of the mean free drift length with increasing resistivity before and after proton irradiation. The small change of the mean free drift length at the highest proton fluence can be understood by considering the high density of introduced defects which shifts the Fermi-level $E_f$ to a value independent of the initial position before irradiation. Assuming that the leakage current density is still a function of $E_f$ position, this also explains the small variation in the leakage current density at the highest proton fluence.

The response to MIPs of the proton irradiated detectors as a function of reverse bias voltage looks similar to the neutron irradiated diodes. There is also an initial region with a linear increase of the signal height and a second region in which the signal only slightly increases with further bias voltages (see e.g. the diodes from wafer MCP90, fig. 24). In analogy to the larger decrease of the CCE both for electrons and holes with proton fluence than with neutron fluence, the signal height after $2*10^{14}$ p/cm² is, for all materials, smaller than at the highest neutron fluence. The signal height at 200 V bias voltage shows the same dependence on the resistivity as the mean free drift length of electrons and decrease from 5250 $e^-$ down to 2500 $e^-$ for the diodes from wafer FR104 (fig. 25). This is due to the fact that the signal primarly is caused by the electron drift in the detector, because of the smaller hole mean free drift length after irradiation.

**4. Comparison of damage functions and determination of the hardness factor**

According to the assumption that the radiation damage scales with the non ionising energy loss (NIEL) of the particles in the detector and the above mentioned production of the same type of defects, it is possible to give the relative damage due to the NIEL of a 1 MeV neutron in GaAs. As mentioned the NIEL of neutrons from the ISIS spallation source is about a factor 2 higher than it would be for a 1 MeV monoenergetic neutron beam. Experimentally we determined the hardness factor for 23 GeV protons, scaling the proton and neutron fluence with the hardness factors to observe a continuous function of the material parameters with the 1 MeV equivalent neutron fluence. Important for this comparison is to look only at one material type, because of the material dependence of the damage functions, like signal height for MIPs and leakage current density. Nevertheless, the determined hardness factor only depends on the particle source and not on the material properties. The hardness factor is also independent of the material parameters. We determined the hardness factor for 23 GeV proton from the PS as $\kappa = 7 \pm 0.9$, which corresponds to a non-ionising energy loss of $NIEL = 21.7 \pm 2.8\ keV/cm$. For example figures 26 and 27 show the signal height for MIPs and the leakage current density as a function of the 1 MeV equivalent neutron fluence.

**5. Summary and conclusion**

We analysed the radiation damage of neutron and proton irradiated detectors made of semi-insulating substrates with resistivities between 0.4 and $8.9*10^7$ Ωcm. A strong dependence between the damage functions and the material was observed. The leakage current density and the breakdown voltage especially can decrease or increase with radiation damage for high or low ohmic materials. However, the leakage current density varies only between 20 and 40 nA/mm² at the highest proton fluence of $2*10^{14}$ $cm^{-2}$. The observed decrease of full active voltage can be explained by the results of C-V measurements performed at low frequencies, which have shown a decrease of net space charge density $N_{net}$ with increasing irradiation. A proposed model including the different ionisation ratios of the introduced deep levels as a function of the resistivities before irradiation, qualitatively explains the decrease of the electron mean free drift length after proton exposure. For all samples at high fluences the mean free drift length of electrons is larger than for holes. This should be considered in a development of strip detectors for high energy experiments, to effectively collect electrons to obtain better radiation hardness. The maximum signal of 5250 $e^-$ for MIPs in a pad detector at a proton fluence of $2*10^{14}$ $cm^{-2}$ was measured with low ohmic material FR12. This is the highest observed value at this fluence and bias voltage of 200 V in a 200 µm thick detector made of semi-insulating GaAs. With a comparison of the radiation damage due to neutrons from the spallation sources at RAL and the protons from the PS at CERN, we determine the non-ionising energy loss of 23 GeV proton as 21.7 ± 2.8 keV/cm.

**Acknowledgements**

We are grateful to M.Edwards (RAL) and F.Lemeilleur (CERN) for the help with irradiation. This work has been supported by the Bundesministerium für Bildung, Wissenschaft, Forschung und Technologie (BMBF) under contract number 05 7FR11I.

## References

RD8 group: Aachen TH I Phys. Inst. Germany, ANSTO Sydney Australia, INFN Italy, Florence Univ. Italy, Freiburg Univ. Germany, Glasgow Univ. Dept. of Physics & Astronomy UK, Lancaster Univ. UK , Modena Univ. Italy, Serpukhov IHEP Russia, DRAL UK, Sheffield Univ. Dept. of Electrical Engineering and Physics UK, Vilnuis, Inst. Phys. Lithuania

[1] S.P. Beaumont, R. Bertin et al., IEEE Trans. Nucl. Sci. NS-40 (4) (1993)

[2] R. Bertin, S. D'Auria et al., Nucl. Instr. And Meth. A 294 (1990) 211

[3] W. Braunschweig, Th. Kubicki et al. , Nucl. Instr. And Meth. A 372 (1996) 111

[4] Technical Proposal of ATLAS Collaboration, CERN/LHCC/94-43, LHCC/P2, 15 December 1994, p. 182

[5] Victor A.J. van Lint, Nucl. Instr. And Meth. A 253 (1987) 453

[6] M. Edwards and D.R.Perry, Rutherford Appleton Laboratory, RAL Report RAL-90-065 (1990)

[7] A.M. Ougouag et al. IEEE Trans. Nucl. Sci. 37 (1990) 2219

[8] A. Van Ginneken, Preprint Fermi National Accelerator Laboratory, FN-522 (1989)

[9] E. León-Florián, C. Leroy and C. Furetta, Particle fluence measurements by activation technique for radiation damage CERN-ECP/95-15 (1995)

[10] H. Norde, J. Appl. Phys. 50 (7) (1979) 5052

[11] F. Dubecky et al. , Semicond. Sci. Technol. 9 (1994) 1654

[12] J.W. Chen et al., Nucl. Instr. and Meth. A 365 (1995) 273

[13] M. Rogalla et al., Nucl. Instr. and Meth. A 380 (1996) 14

[14] M. Rogalla et al., Influence of the compensation in semi-insulating GaAs on the particle detector performance, Proc. SIMC-9, Toulouse, 1996 (IEEE)

[15] G. Cavalier et al., Nucl. Instr. and Meth. 92 (1971) 137

[16] J.F. Ziegler, J.B. Biersack, Transport of Ions in Matter (Trim Version 03)

[17] F.Lemeilleur et al., Nucl. Instr. and Meth. A 360 (1995) 438

[18] E.A. Burke et al., IEEE Trans. Nucl. Sci. Vol. 34 (1987) 1220

[19] F. Dubecky et al., Neutron irradiated undoped SI-GaAs: Galvanomagnetic, I-V, PC and alpha detection studies, Proc. of the 3rd int. Workshop on GaAs and related compounds, S.Minato, Italy (1995), Word scientific

[20] M.Rogalla et al., Analysis of trapping and detrapping in semi-insulating GaAs detectors, submitted to Nucl. Instr. and Meth. A

[21] R.Irsigler et al., Contact and substrate influence on semi-insulating GaAs detectors, submitted to Nucl. Instr. and Meth. A

[22] D.Pons and J.C. Bourgin, Solid State Phys. 18 (1985) 3839

[23] Victor A.J. van Lint, Nucl. Instr. and Meth. A 253 (1987) 453

[24] J. Krueger, et al., The influence of native point defects on the performance of diodes built on semi-insulating GaAs, Proc. SIMC-9, Toulouse, 1996 (IEEE)

| substrates | contacts* | guard ring | diameter [mm] | resistivity [$10^7$ Ωcm] 20°C | radiation source |
|---|---|---|---|---|---|
| FR12 | SS: Ti/Al | yes | 3 | 0.42 | p |
| OUT44 | SO: TiN/Ti-W Ni/Ge/Au/Ni/Au | yes | 2 | 2.1 | p |
| OUT50 | SS: Ti/Ti-W | yes | 2 | 2.4 | p |
| FR41 | SO: Ti/Pt/Au-Ni/Ge/Au/Ni/Au | no | 3 | 4.8 | n, p |
| MCP90 | SO: Ti/Pt/Au-Ni/Ge/Au/Ni/Au | no | 3 | 4.9 | n, p |
| FR76 | SO: Ti/Pt/Au-Ni/Ge/Au/Ni/Au | no | 3 | 5.9 | n, p |
| FR104 | SS: Ti/Ti-W | yes | 3 | 8.9 | p |

Table 1: Parameter of the irradiated diodes.

* SO: Schottky - Ohm, SS: Schottky - Schottky

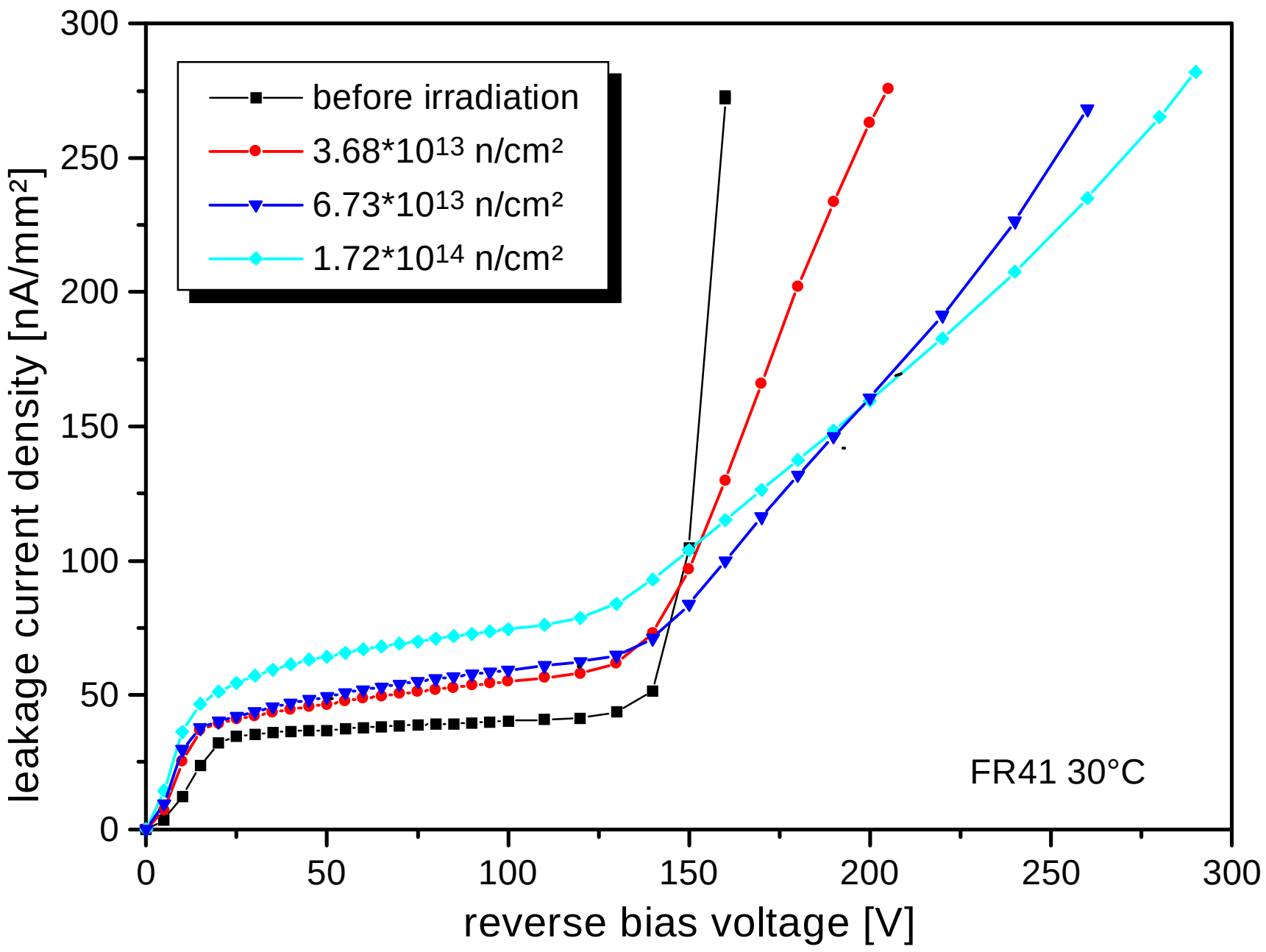

Figure 1: The leakage current density (for example of diodes on wafer FR41, 30°C) versus the reverse bias voltage for different neutron fluences.

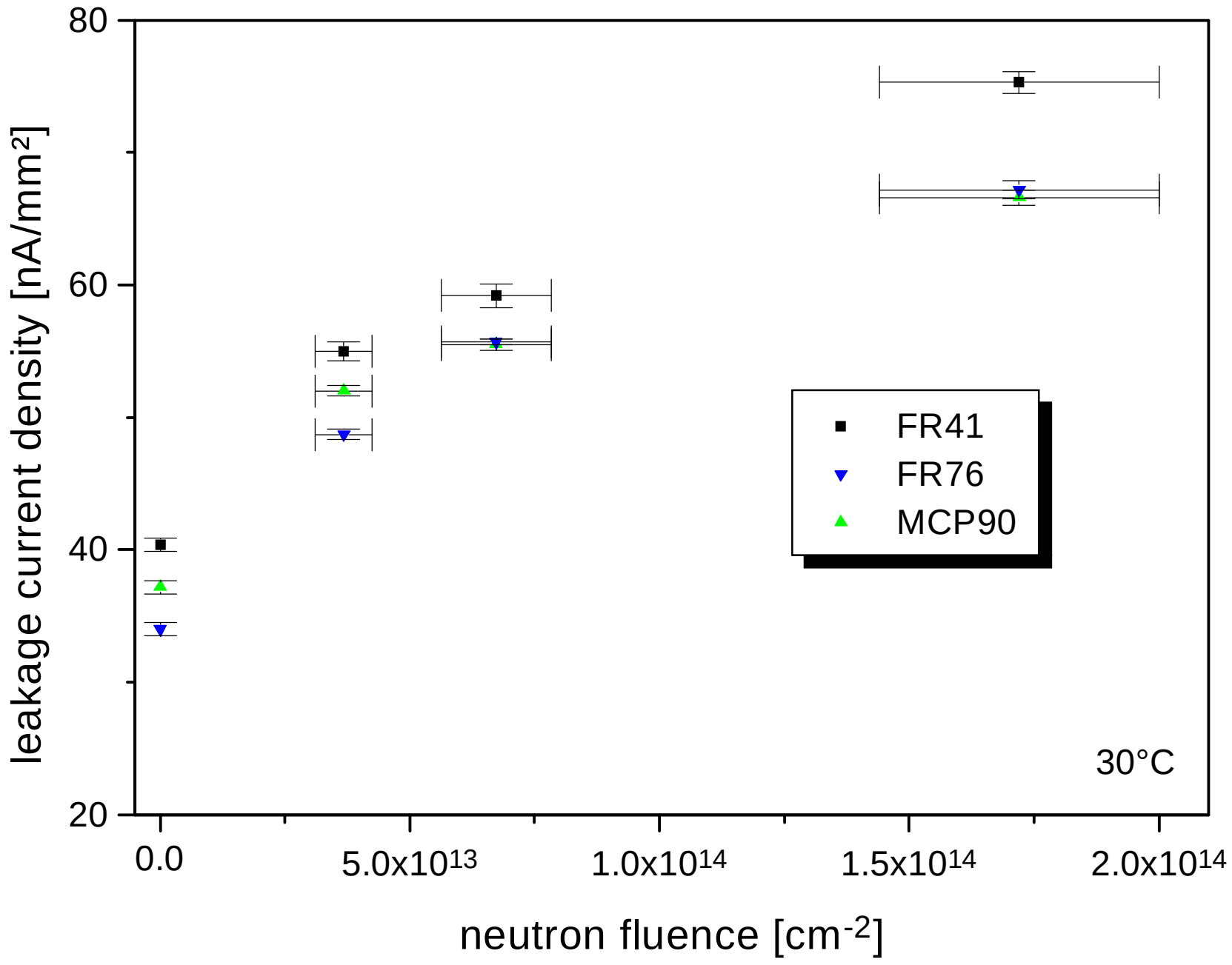

Figure 2: The leakage current density at a bias voltage of 100 V and 30°C of the different detectors as a function of neutron fluence.

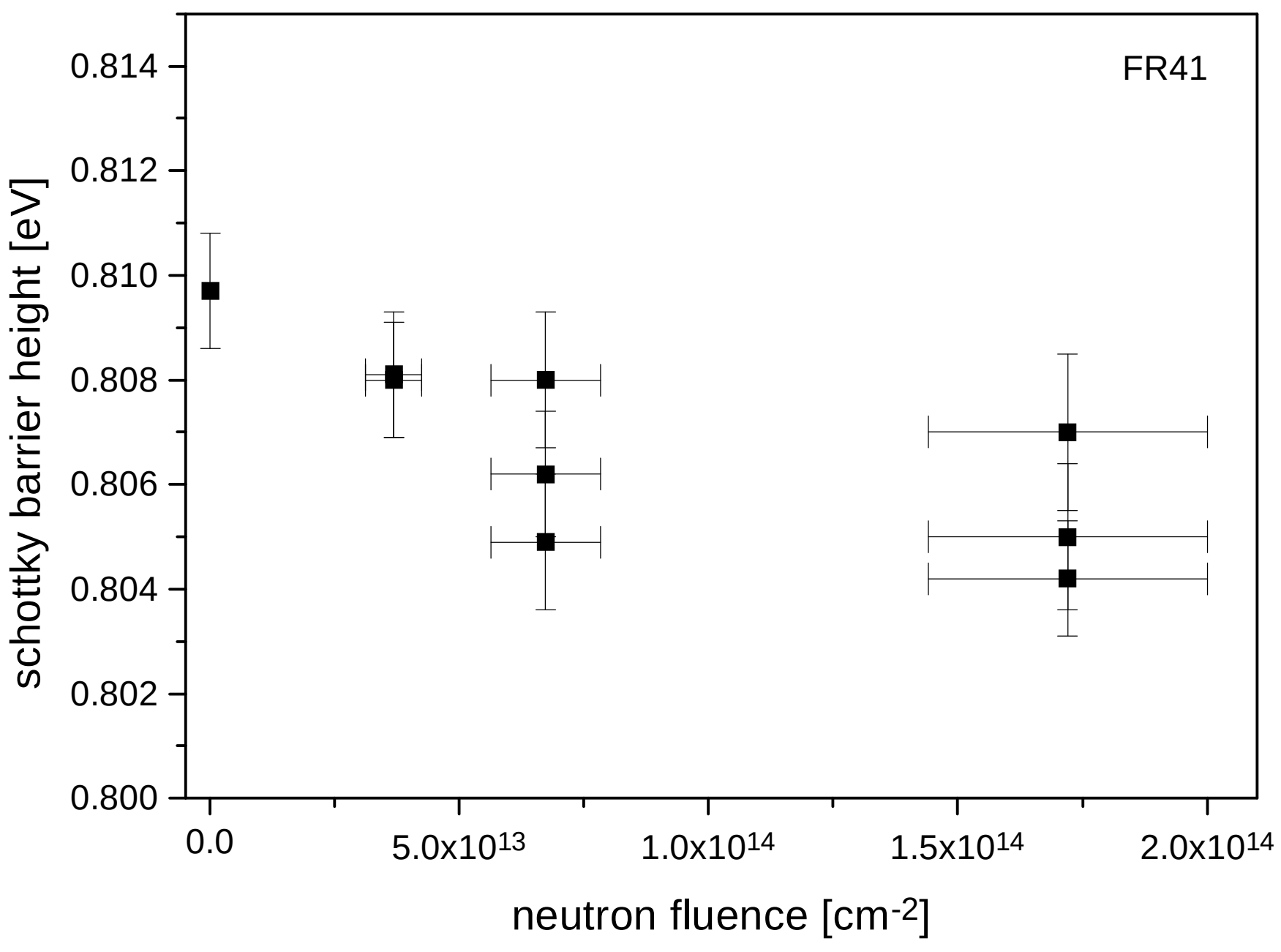

Figure 3: The Schottky barrier height determined from the Norde plot as a function of radiation level (diodes from wafer FR41).

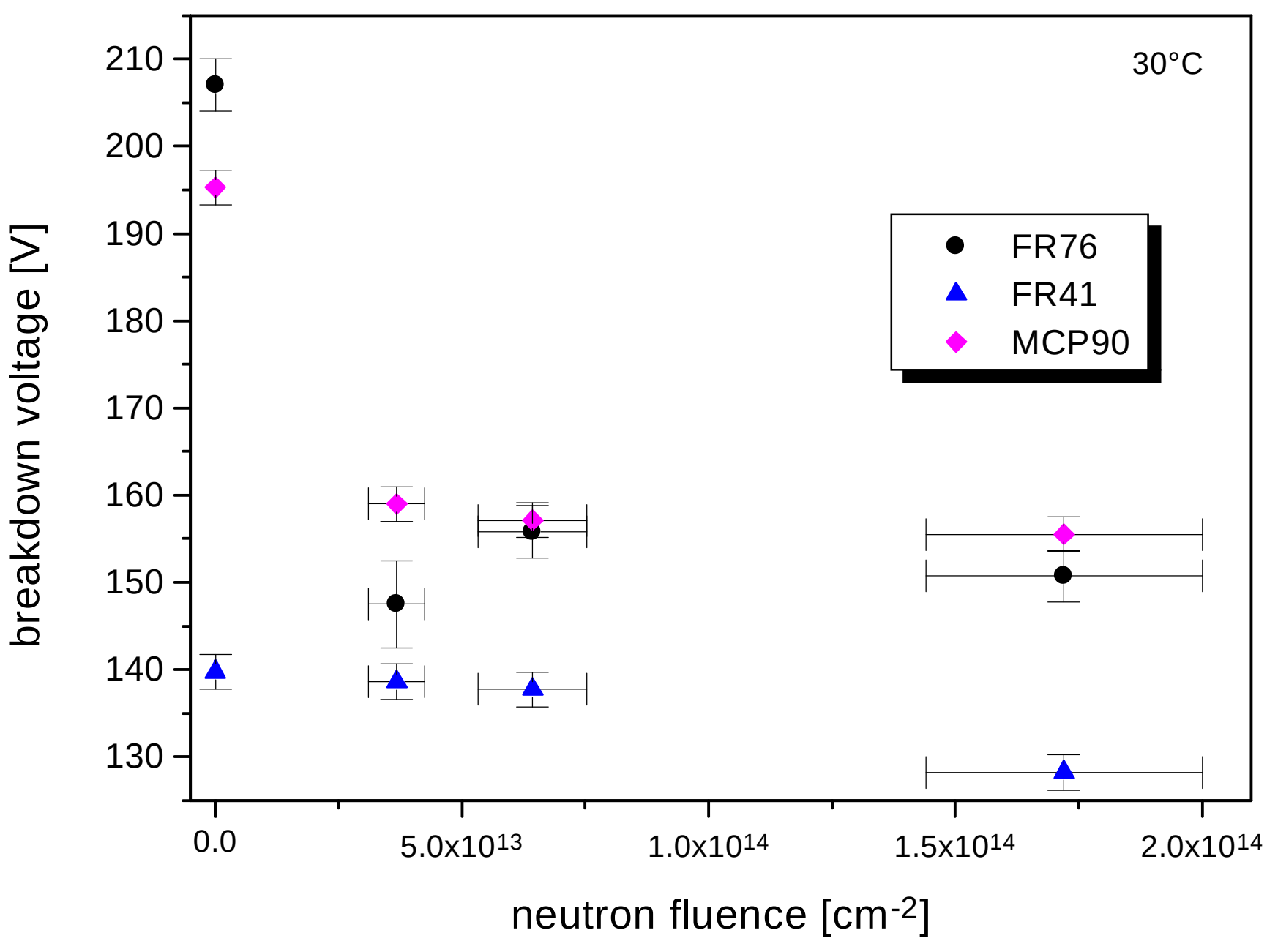

Figure 4: The breakdown voltage of the detectors as a function of neutron fluence (30°C).

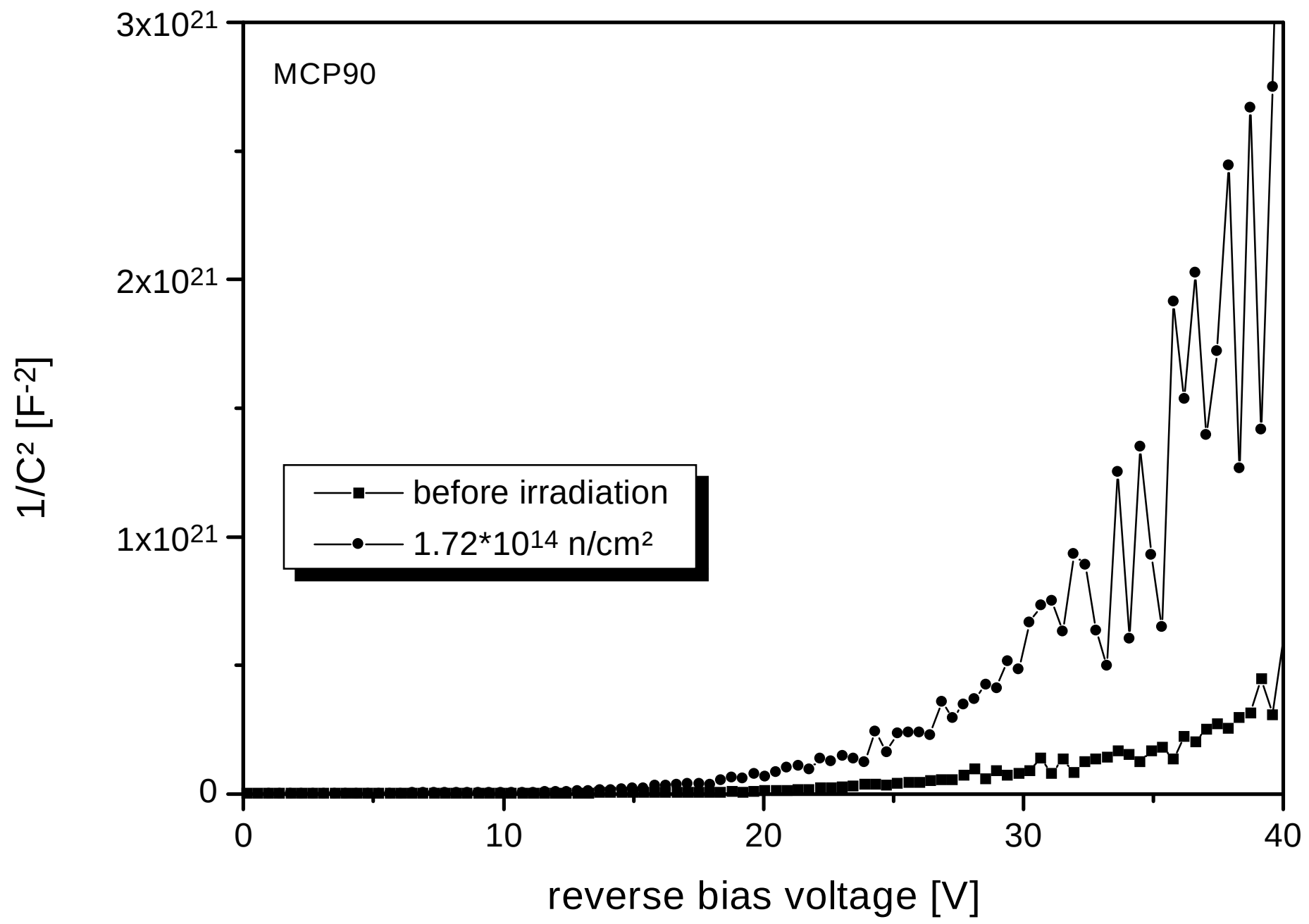

Figure 5: Typical $1/C^2$-V plot for SI-GaAs Schottky diodes measured at room temperature and with 10 Hz before and after irradiation with a fluence of $1.72*10^{14}$ n/cm² (MCP90).

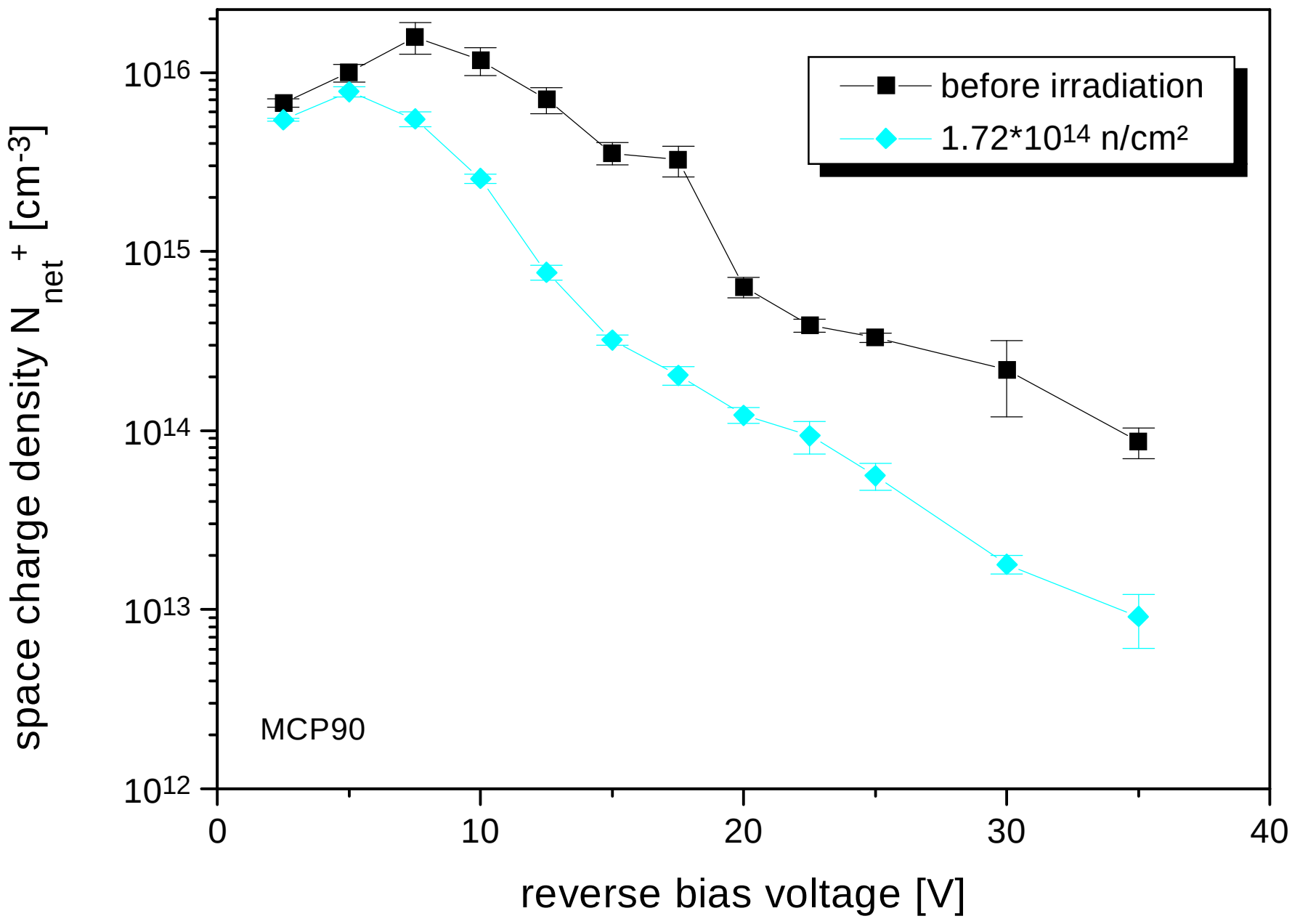

Figure 6: Space charge density calculated from the $1/C^2$-plot in figure 5.

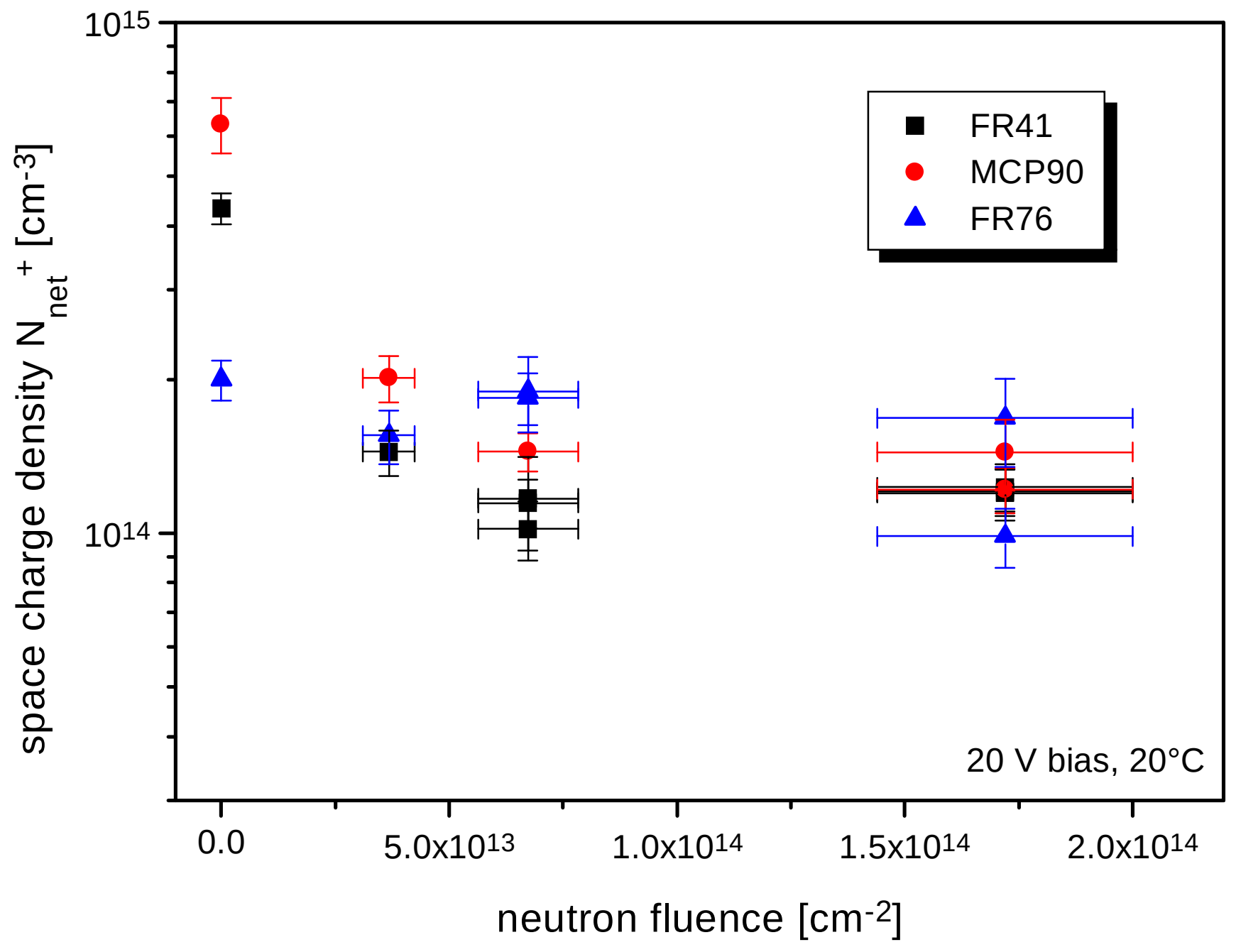

Figure 7: Space charge density at 20 V bias voltage as a function of neutron fluence for the three different materials.

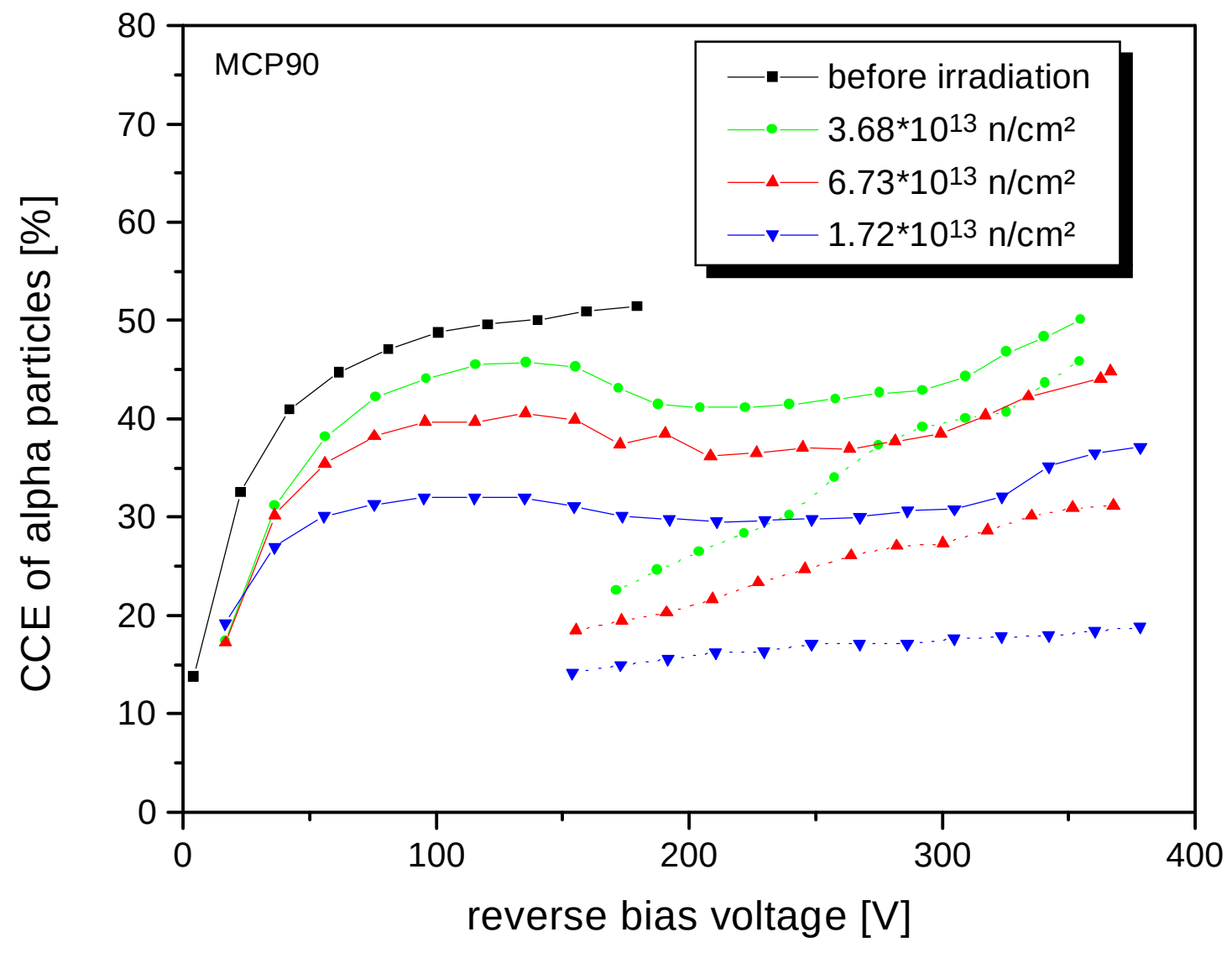

Figure 8: Example of the CCE of α-particles of detectors from wafer MCP90 versus the reverse bias voltage before and after irradiation with neutrons. The solid lines give the values for front side exposure and the dotted lines shows the values for back side irradiation with α-particles.

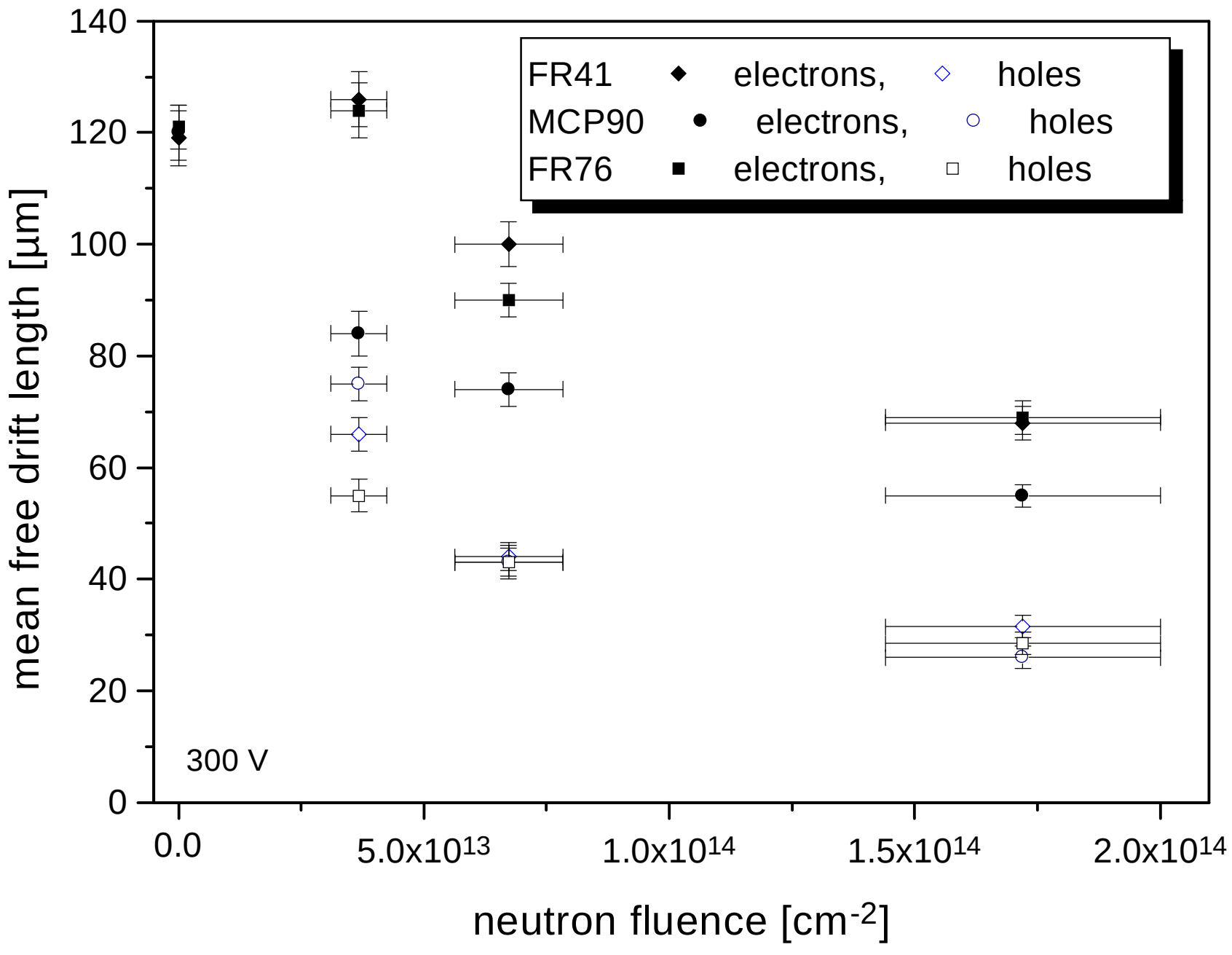

Figure 9: The mean free drift length for both carriers calculated from the CCE of α-particles versus the neutron fluence (bias voltage 300 V).

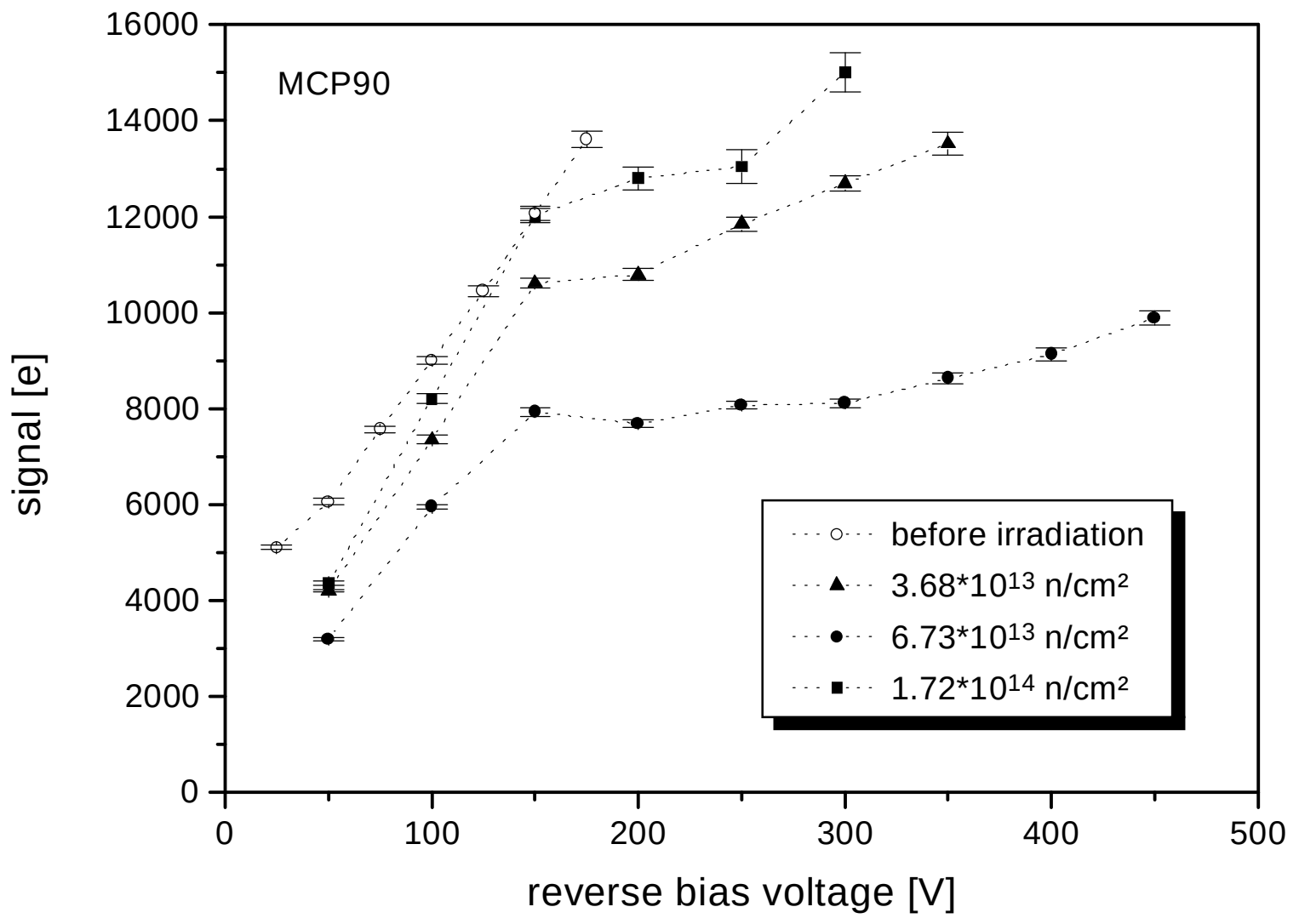

Figure 10: The most probable value of signal height for $^{90}$Sr electrons (shaping time 500 ns) as a function of reverse bias voltage before and after neutron irradiation with different fluences for wafer MCP90.

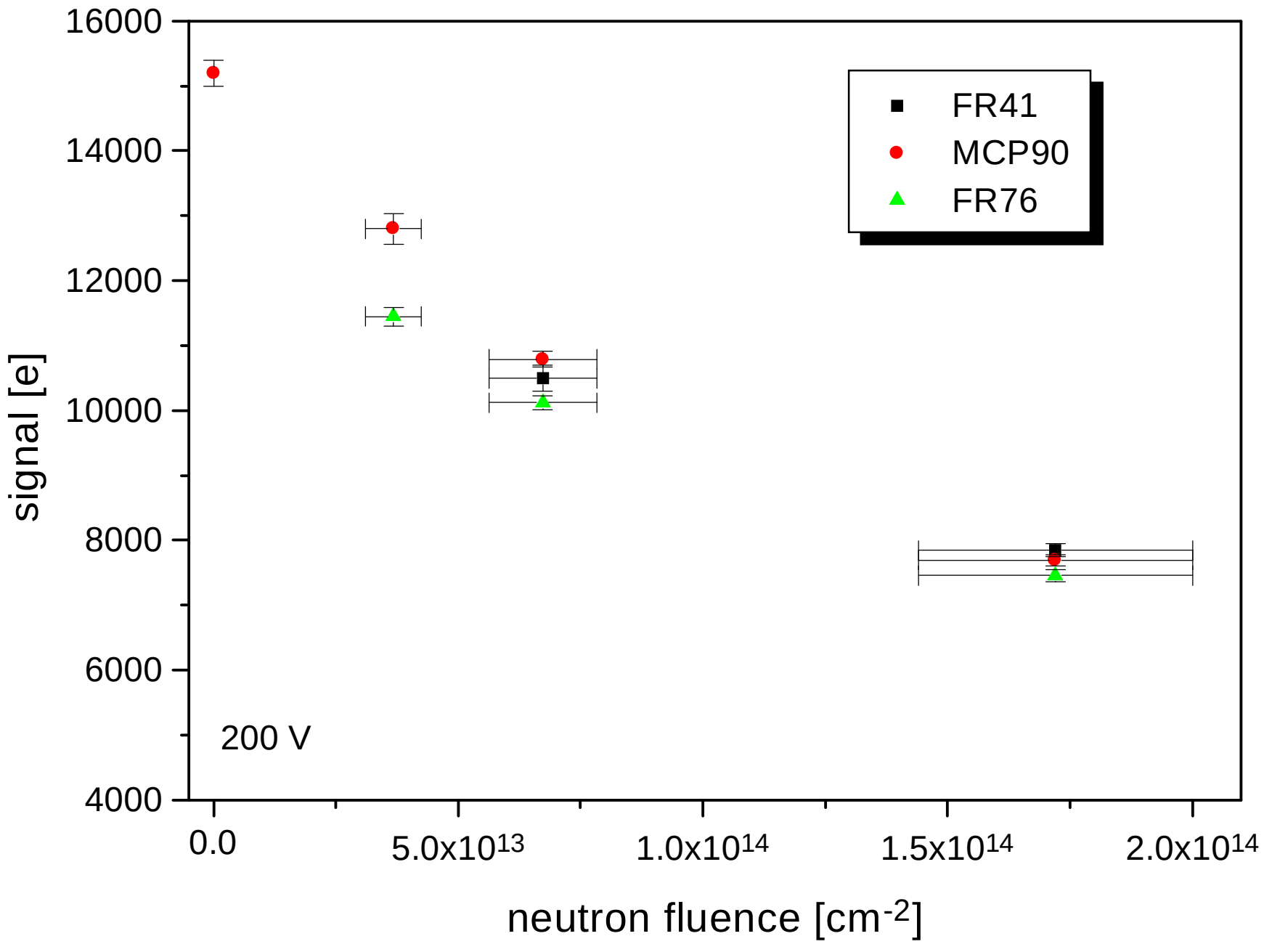

Figure 11: The signal pulse height data for all materials as a function of neutron fluence at a reverse bias voltage of 200 V.

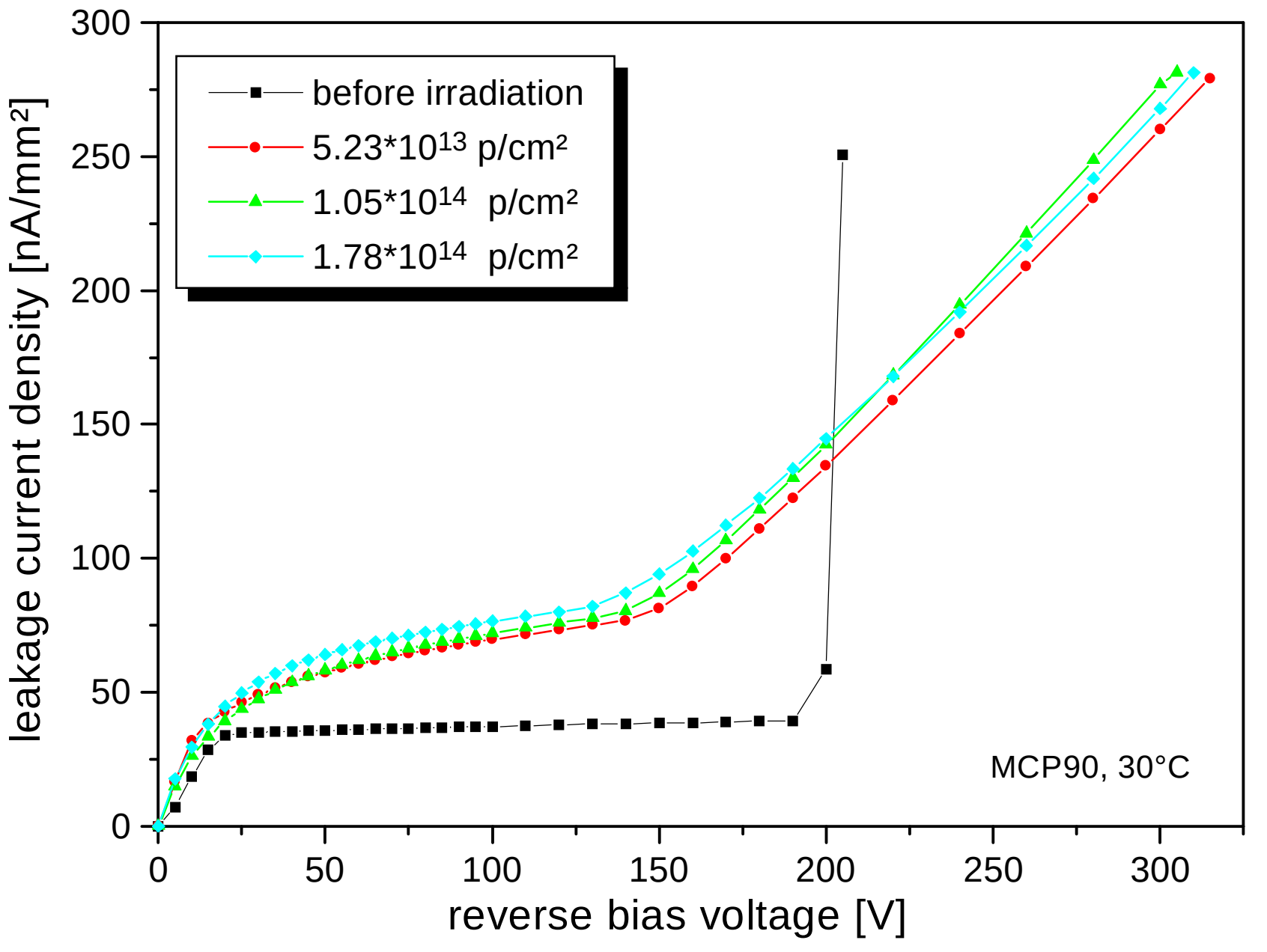

Figure 12: The leakage current density of diodes from the wafer MCP90 at 30°C versus the reverse bias voltage for different proton fluences.

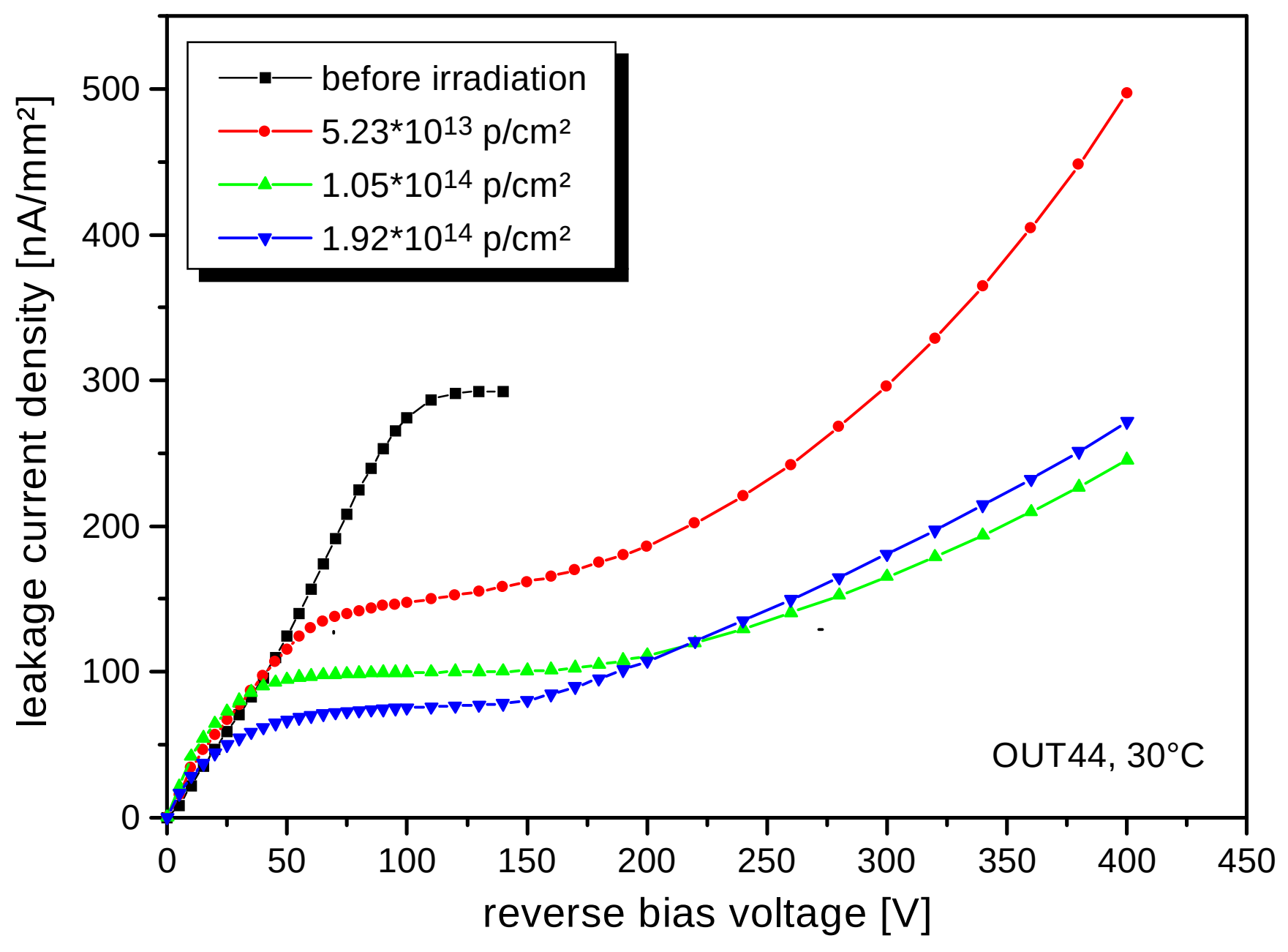

Figure 13: The leakage current density of diodes from the low-ohmic wafer OUT44 at 30°C versus the reverse bias voltage for different proton fluences.

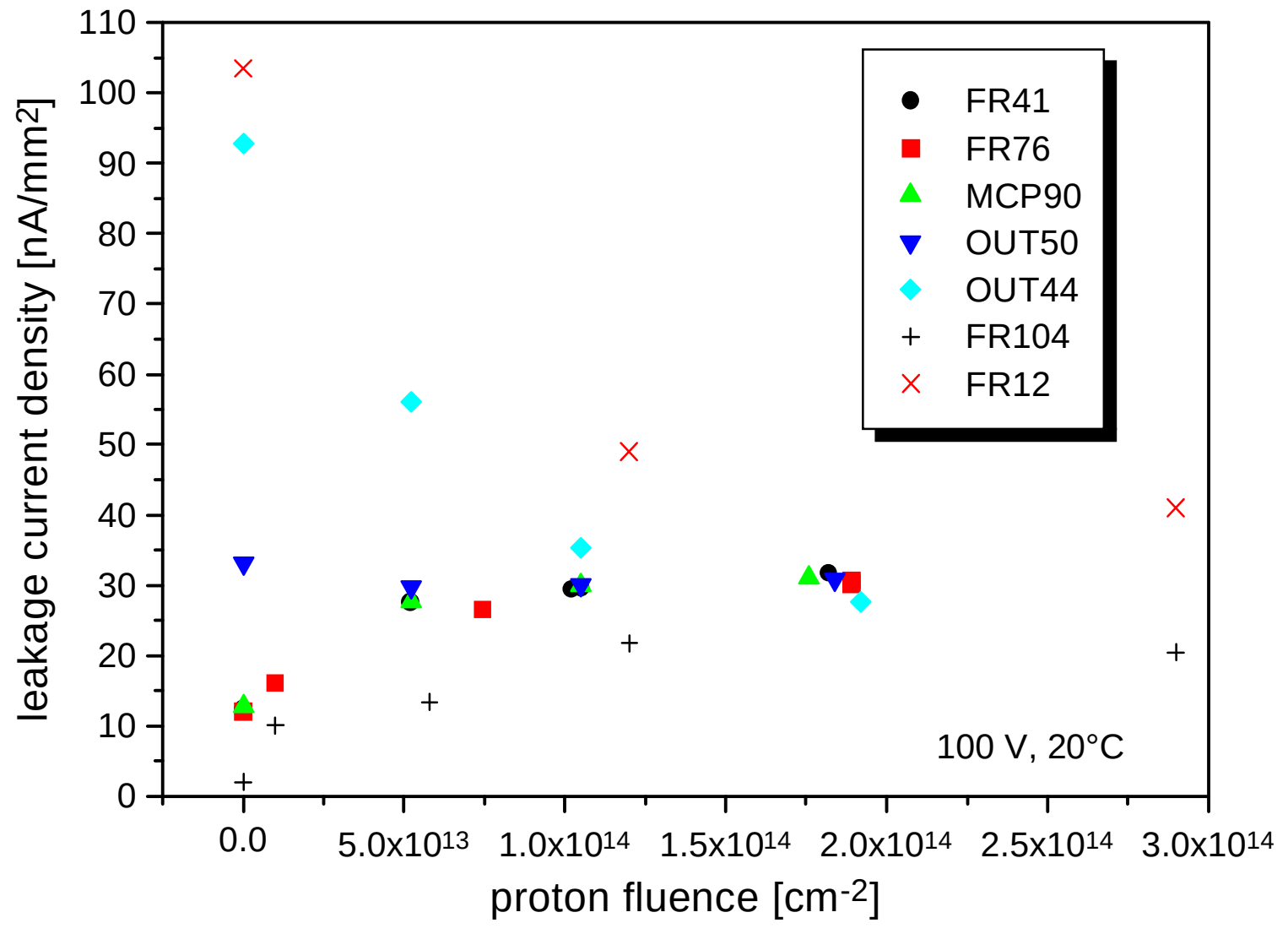

Figure 14: The leakage current density at a bias voltage of 100 V and 30°C of the different detectors as a function of proton fluence.

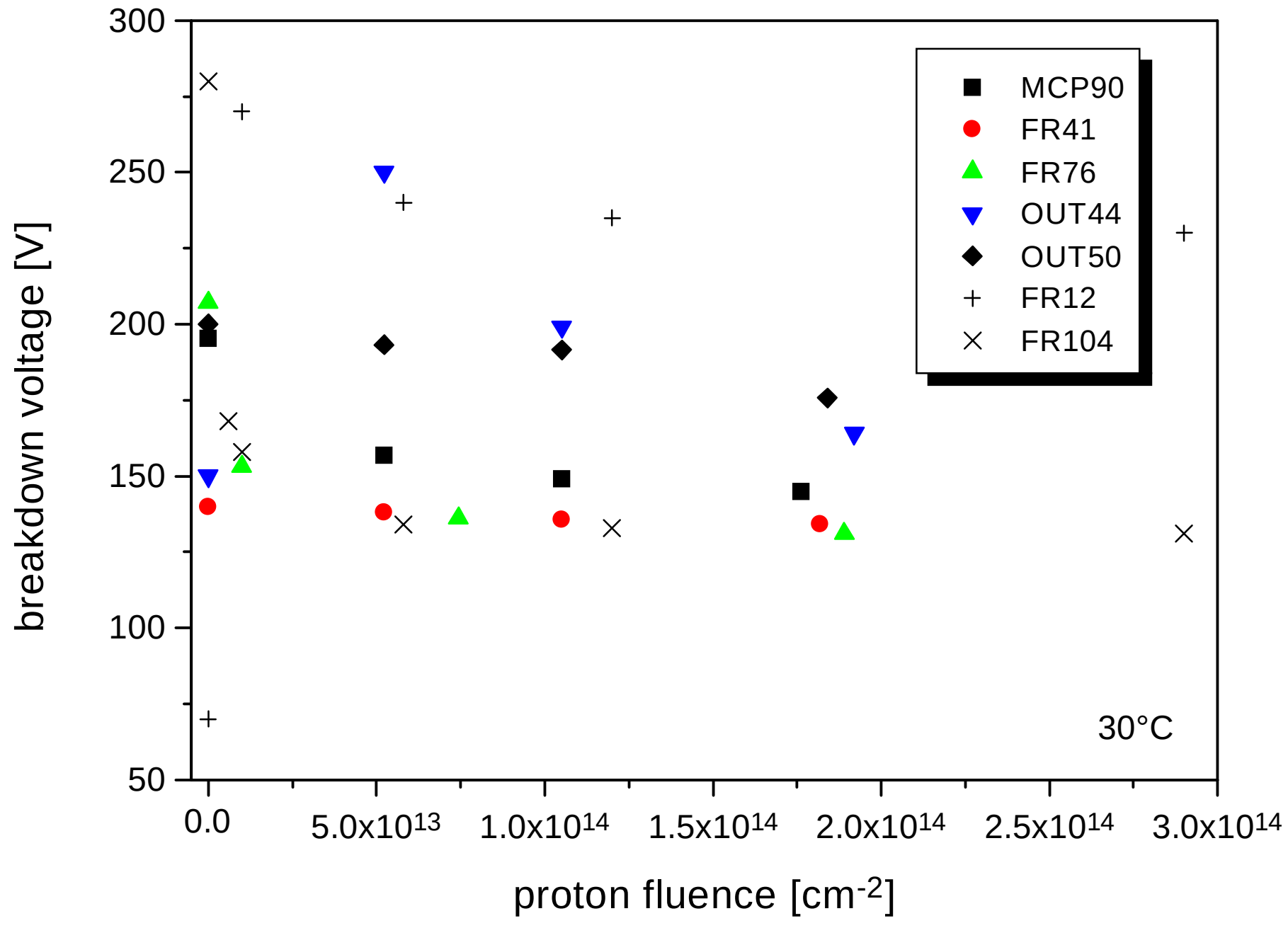

Figure 15: The breakdown voltage of the detectors as a function of proton fluence (30°C).

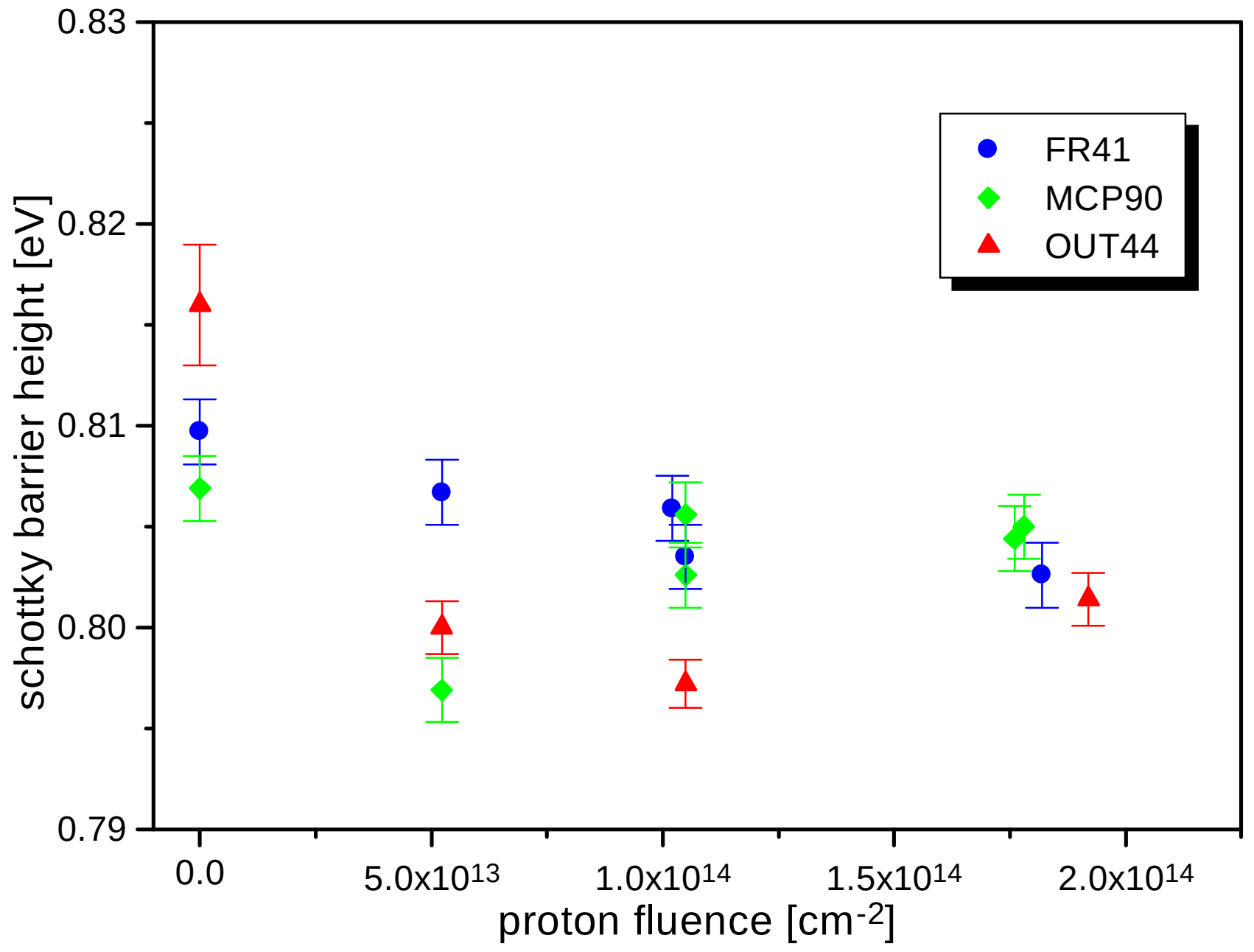

Figure 16: The Schottky barrier height determined from the Norde plot of the detectors with ohmic back contacts as a function of proton radiation level.

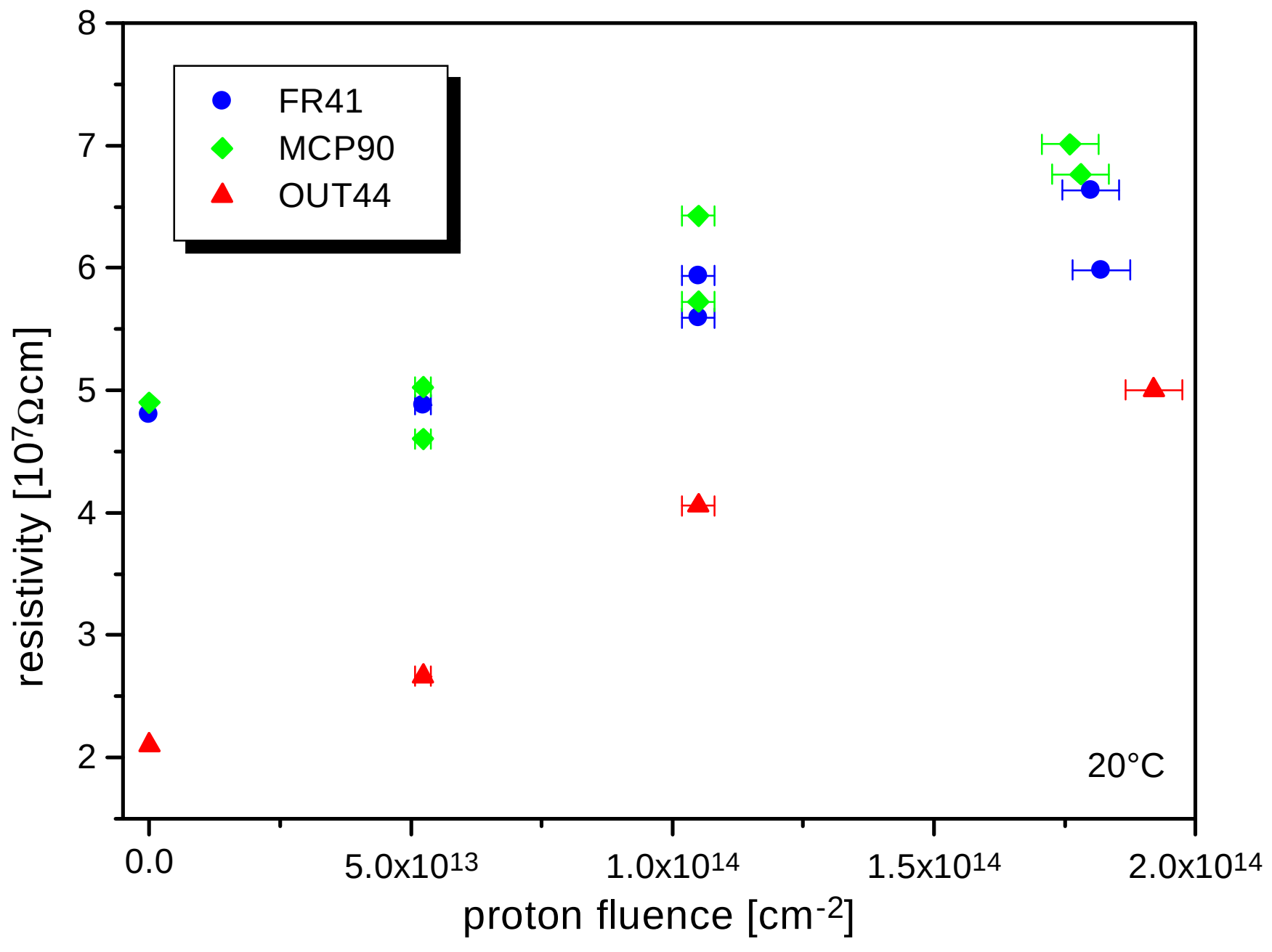

Figure 17: The substrate resistivity at 20°C determined from the Norde plot of the detectors with ohmic back contacts as a function of proton fluence.

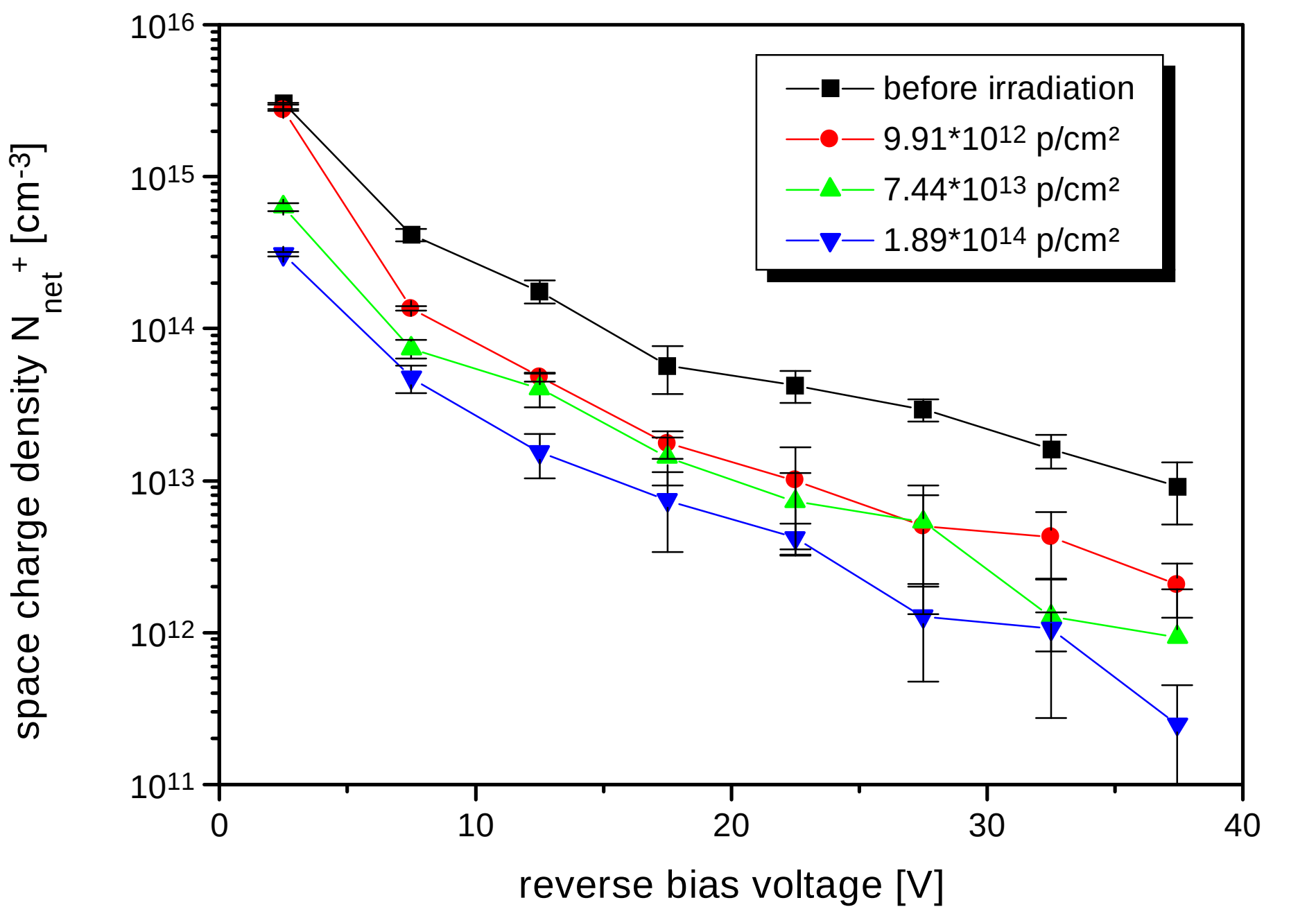

Figure 18: The net space charge density as a function of reverse bias voltage and proton fluence determined with C-V measurements for the diodes from wafer FR76.

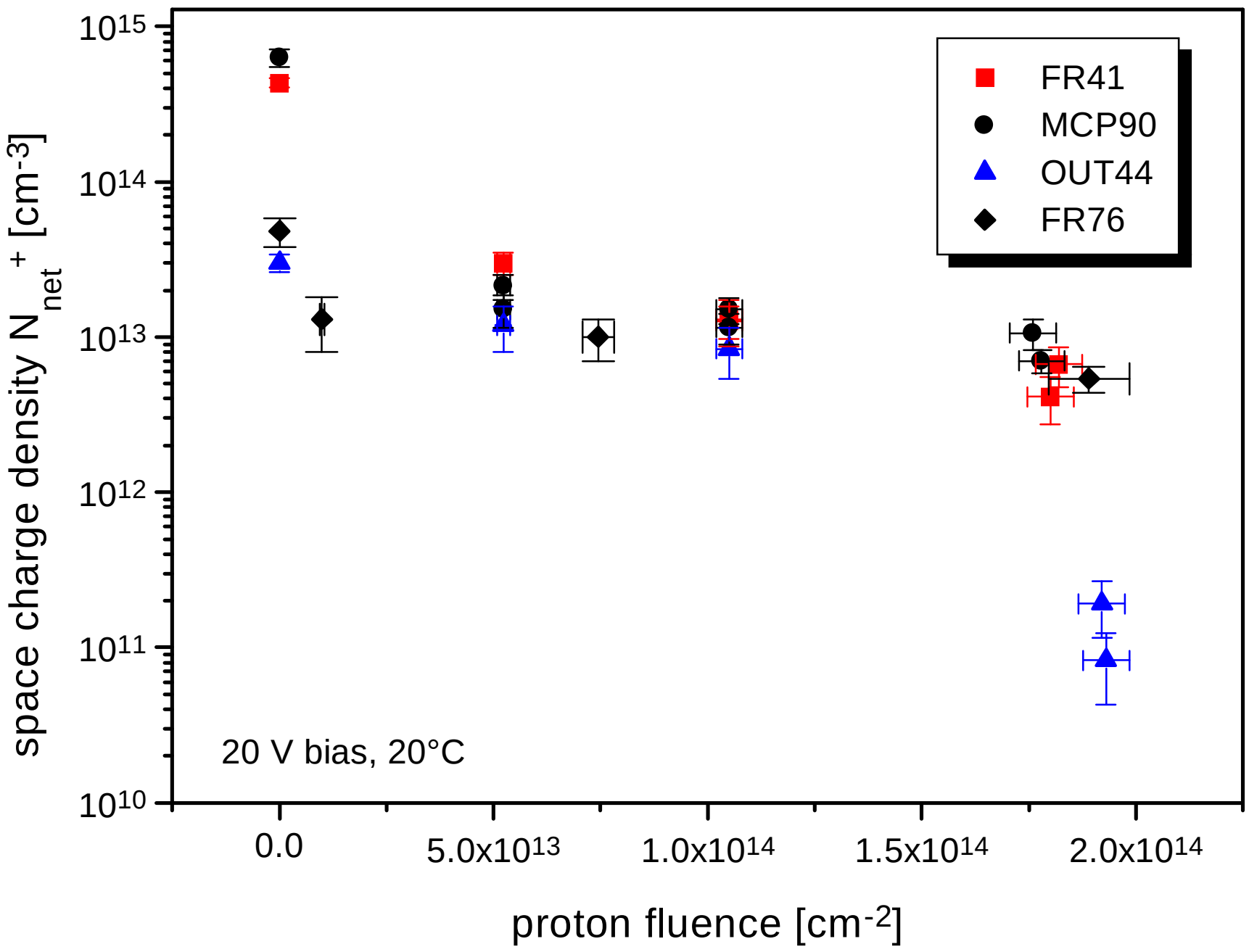

Figure 19: The net space charge density at 20 V reverse bias voltage as a function of proton fluence for four different materials.

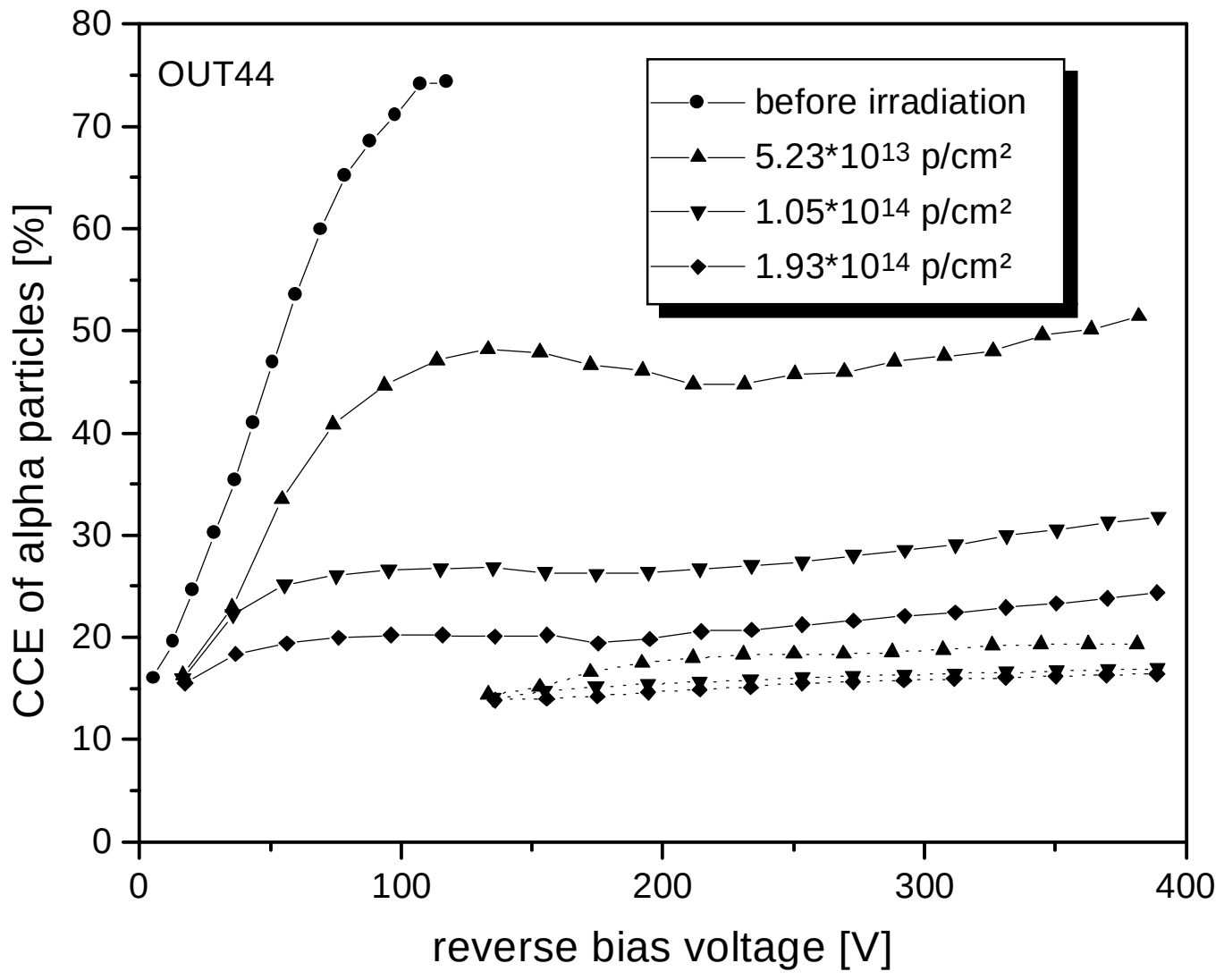

Figure 20: Example of the CCE of α-particles of detectors from wafer OUT44 versus the reverse bias voltage before and after irradiation with protons. The solid lines give the values for front side exposure and the dotted lines shows the values for back side irradiation with α-particles.

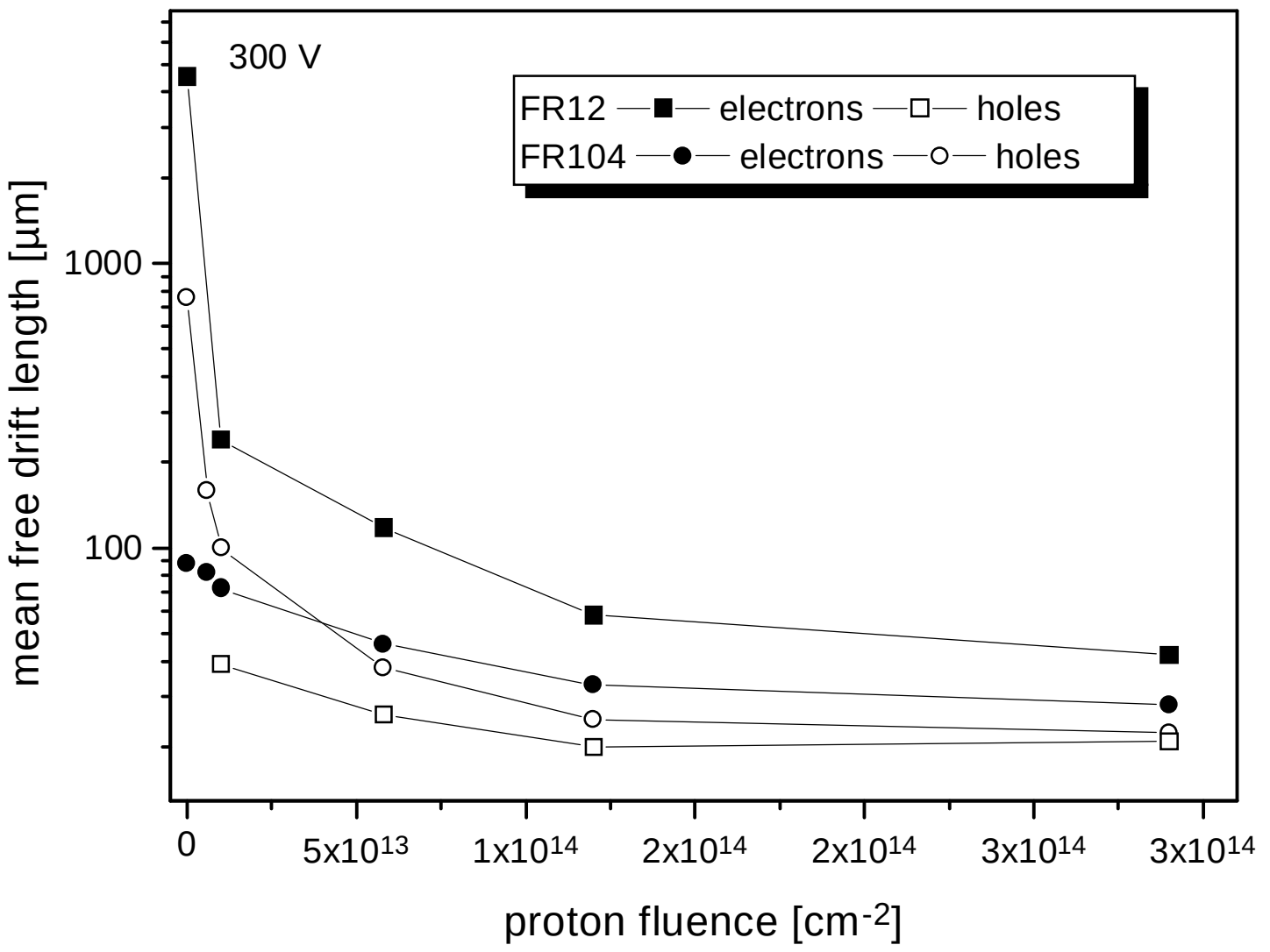

Figure 21: The mean free drift length of the charge carriers in the materials FR12 and FR104 as a function of proton fluence determined from the CCE measurements at 300 V reverse bias voltage.

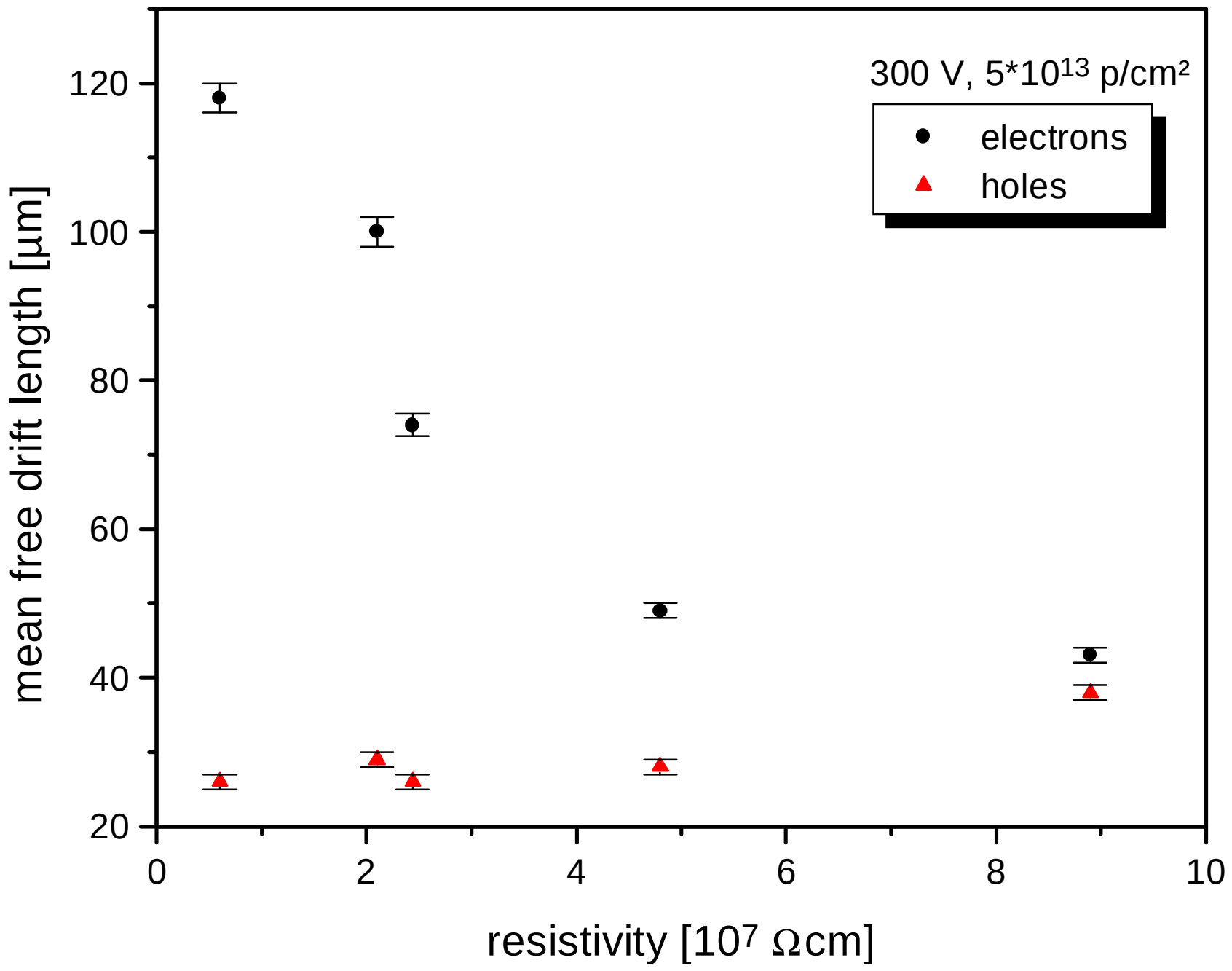

Figure 22: The mean free drift length of electrons and holes at 300 V bias and a proton fluence of $5*10^{13}$ $cm^{-2}$ as a function of resistivity before irradiation.

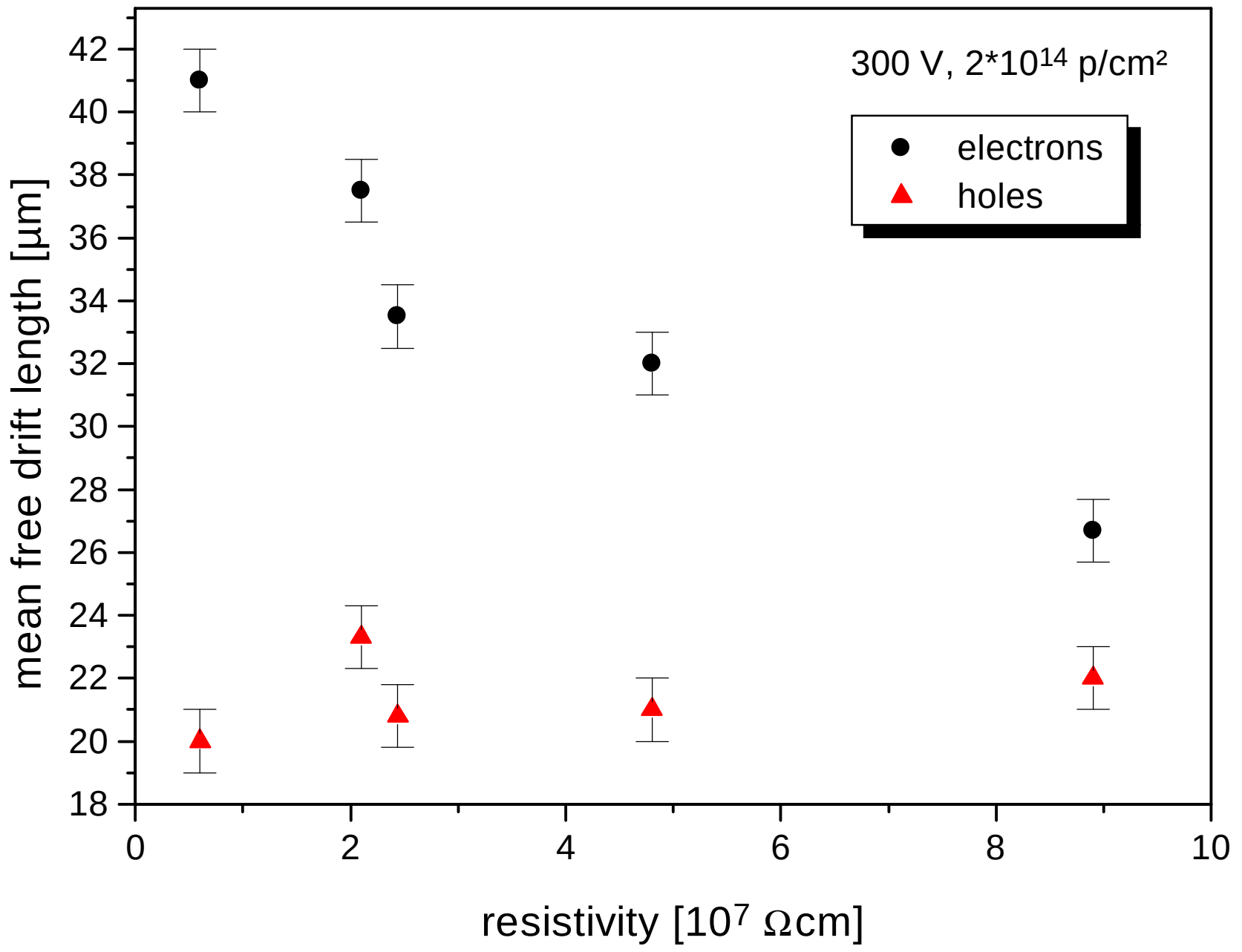

Figure 23: The mean free drift length of electrons and holes at 300 V bias and a proton fluence of $2*10^{14}$ $cm^{-2}$ as a function of resistivity before proton irradiation.

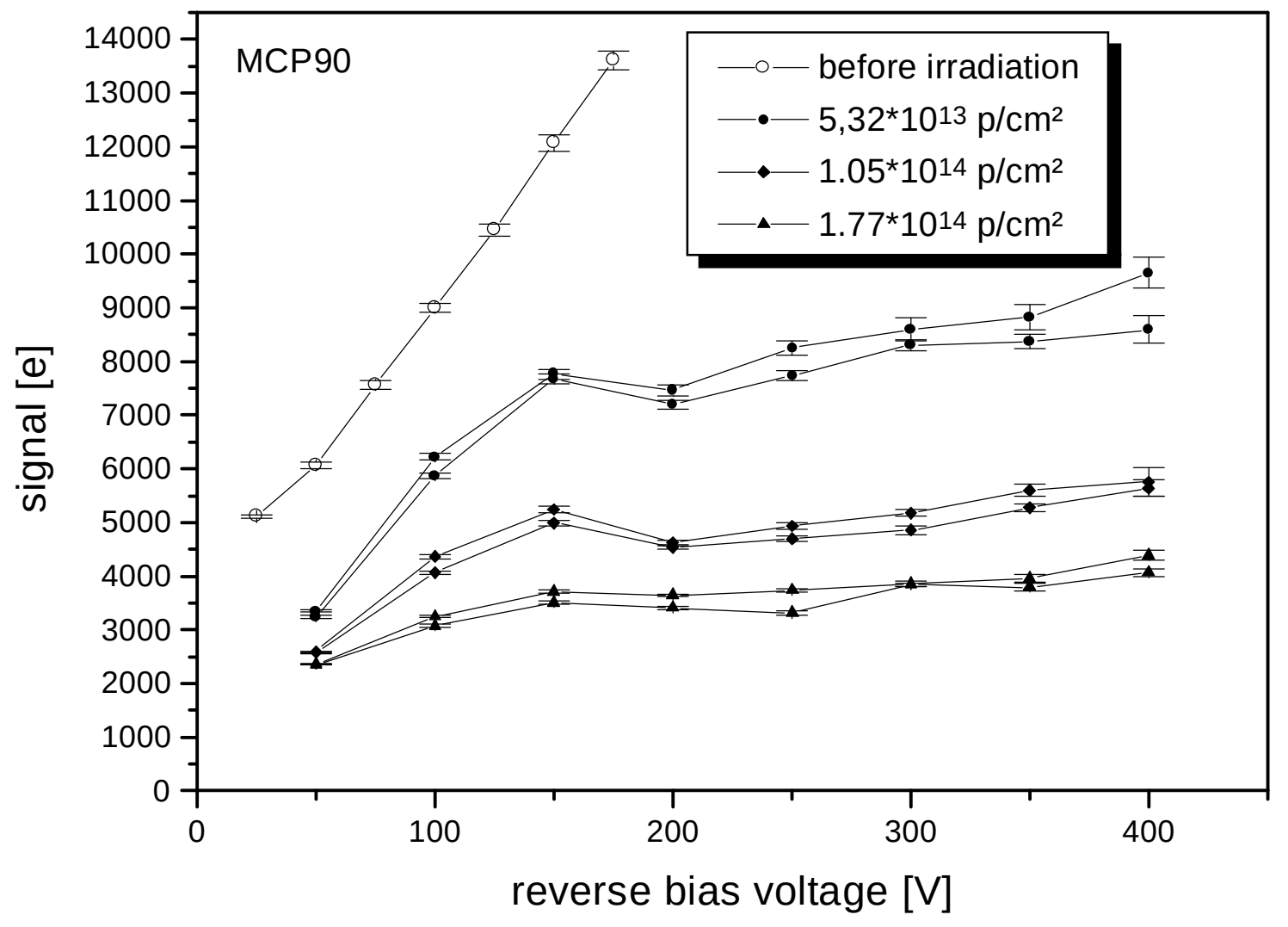

Figure 24: The most probable value of signal height for $^{90}Sr$ electrons (shaping time 500 ns) as a function of reverse bias voltage before and after proton irradiation with different fluences for the diodes from wafer MCP90.

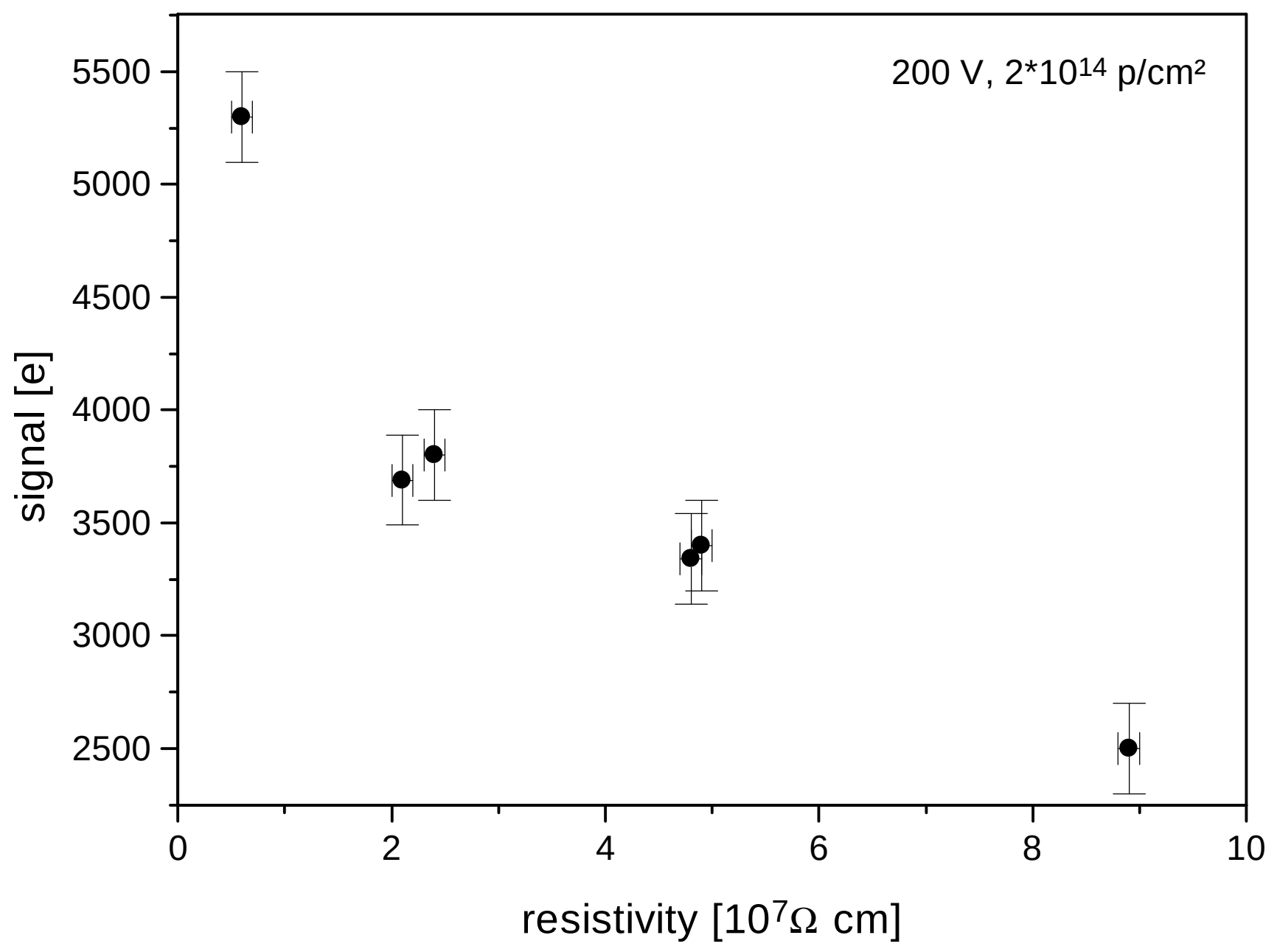

Figure 25: The most probable value of signal height for $^{90}Sr$ electrons as a function of the substrate resistivity before irradiation.

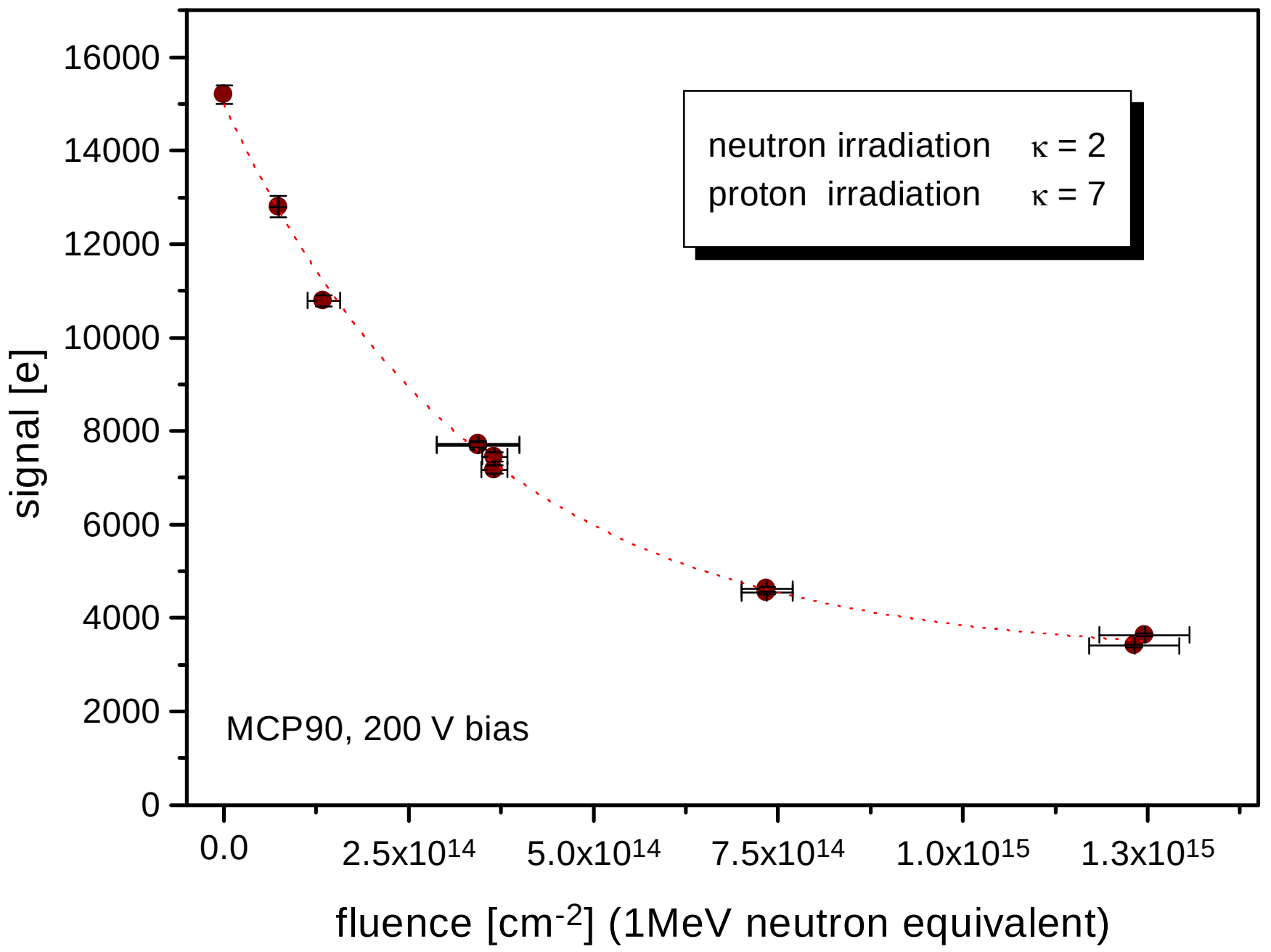

Figure 26: The signal response due to MIPs versus the 1 MeV neutron equivalent fluence of the detectors from wafer MCP90 at 200 V bias.

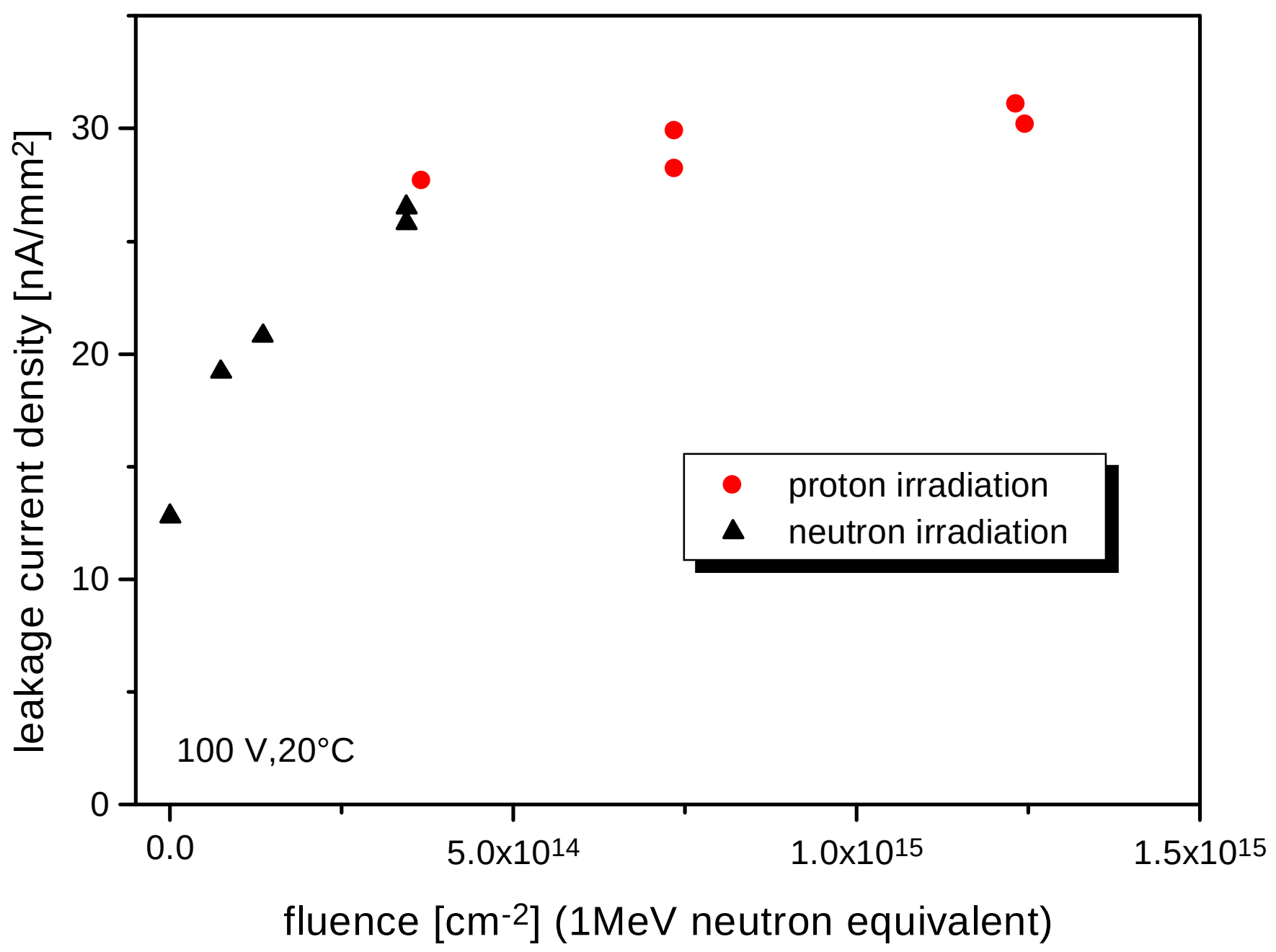

Figure 27: The leakage current density as a function of the 1 MeV neutron equivalent fluence, determined in the neutron and proton irradiated diodes from wafer MCP90 at 100 V bias and 20°C.